\documentclass[a4paper,11pt]{article}
\usepackage{jheppub} 
\usepackage{lineno}
\usepackage{float}
\usepackage[dvipsnames]{xcolor}
\usepackage{orcidlink}
\usepackage{siunitx}  
\usepackage{verbatim}
\usepackage{comment}
\usepackage{multirow}
\usepackage{multicol}

\title{\boldmath New constraints on physics within and beyond the standard model from the latest CONUS datasets}

\author[1]{\orcidlink{0009-0000-2122-5186}N.~Ackermann,}
\author[1]{H.~Bonet,}
\author[1, a]{\orcidlink{0000-0002-0218-2835}A.~Bonhomme,}
\author[1]{\orcidlink{0000-0002-5751-5289}C.~Buck,}
\author[2]{K.~F\"ulber,}
\author[1, b]{\orcidlink{0000-0003-0470-3320}J.~Hakenm\"uller,}
\author[1]{J.~Hempfling,}
\author[1]{G.~Heusser,}
\author[1]{\orcidlink{0000-0001-9788-4014}T.~Hugle,}
\author[1]{\orcidlink{0000-0002-3704-6016}M.~Lindner,}
\author[1,d]{\orcidlink{0000-0003-0320-7827}W.~Maneschg,}
\author[1]{\orcidlink{0000-0002-7280-0854}S.~Mertens,}
\author[1]{K.~Ni,}
\author[1]{\orcidlink{0009-0006-5637-4441}D.~Piani,}
\author[3]{M.~Rank,}
\author[4]{\orcidlink{0000-0002-9293-1106}T.~Rink,}
\author[1]{\orcidlink{0000-0001-8014-4079}E.~Sánchez García,}
\author[3]{I.~Stalder,}
\author[1]{H.~Strecker,}
\author[2]{R.~Wink,}
\author[3,c]{and J.~Woenckhaus}

\affiliation[1]{Max-Planck-Institut f\"ur Kernphysik, Saupfercheckweg 1, 69117 Heidelberg, Germany}
\affiliation[2]{PreussenElektra GmbH, Kernkraftwerk Brokdorf, Osterende,
25576 Brokdorf, Germany}
\affiliation[3]{Kernkraftwerk Leibstadt AG, 5325 Leibstadt, Switzerland}
\affiliation[4]{Institut f\"ur Astroteilchenphysik, Karlsruher Institut f\"ur Technologie (KIT),
Hermann-von-Helmholtz-Platz 1, 76344 Eggenstein-Leopoldshafen, Germany}

\affiliation[a]{\textit{Present address:} University of Strasbourg, CNRS, IPHC UMR 7178, F-67000 Strasbourg, France}
\affiliation[b]{\textit{Present address:} Marietta Blau Institute  for Particle Physics of the Austrian Academy of Sciences (MBI), Dominikanerbastei 16, 1010 Vienna, Austria}
\affiliation[c]{\textit{Present address:} Paul Scherrer Institut, Forschungsstrasse 111, 5232 Villigen, Switzerland.}
\affiliation[d]{\textit{Present address:} Mirion Technologies (Canberra) GmbH, Stahlstraße 42–44, 65428 Rüsselsheim, Germany}

\collaboration{CONUS Collaboration}
\emailAdd{conus.eb@mpi-hd.mpg.de}

\date{\today}

\abstract{
\noindent
Its detections with pion-decay-at-rest, solar and recently with reactor antineutrinos by the \textsc{Conus} collaboration render coherent elastic neutrino-nucleus scattering (CE$\nu$NS) an established tool for investigations within and beyond the Standard Model (SM).
The \textsc{Conus} experiment located at the nuclear power plants in Brokdorf (Germany) and Leibstadt (Switzerland) operates Germanium semiconductor detectors in a compact shield at close distance to the reactor core. An observation with $3.7 \sigma$ significance is reported at the Leibstadt site, showing good agreement with its SM prediction.
Physics investigations performed with the last datasets collected at the Brokdorf reactor and with the first data obtained at the Leibstadt site are summarized. By using the experimental analysis framework, the presented results contain the full systematics that underlie the experiment.
Previously determined limits with neutrino-electron scattering on the neutrino magnetic moment and a neutrino millicharge are improved to $\mu_{\nu} <5.18\cdot  10^{-11}\mu_\mathrm{B}$ and $q_{\nu}<1.76\cdot  10^{-12} e_0$ (90\% C.L).
Further, the scale of new physics related to NSIs is improved to $\Lambda_{\rm NSI}$=145\,GeV and limits on the coupling of light new mediators are lowered down to $4 \cdot 10^{-7}$ (90\% C.L.) with the new data.
Finally, the determination of the Weinberg angle with CE$\nu$NS and reactor antineutrinos yields $\sin^{2}\theta_W= 0.28^{+0.03}_{-0.04}$ at a momentum transfer of $\sim 10 \ \mathrm{MeV}$.
}

\begin{document}
\maketitle
\flushbottom


\section{Introduction}
\label{sec:intro}

With the detections of coherent elastic neutrino-nucleus scattering (CE$\nu$NS) in recent years, new opportunities have emerged for investigations within and beyond the standard model (SM).
While already predicted in the 1970s as a weak, neutral current interaction channel~\cite{Freedman:1973yd,Kopeliovich:1974mv,Tubbs:1975jx}, it remained undetected for decades due to its low energy properties.
In CE$\nu$NS a neutrino with $\mathcal{O}$(MeV) energies elastically scatters off a nucleus leaving a small nuclear recoil as imprint in the detector medium.
Although the energy configuration of the neutrino allows for a coherent sum of individual scattering amplitudes of the nucleus' constituents and, thus, a high scattering probability, the resulting nuclear recoil energy is still tiny due to the mass differences between the involved particles.
Nevertheless, CE$\nu$NS is predicted to be the strongest interaction channel of neutrinos in the SM - orders of magnitude stronger than inverse beta decay - when it is observable, i.e.\  low-energy nuclear recoils become experimentally accessible. 
As experimental progress continued over the decades this has finally been achieved: first with (anti-)neutrinos from a pion-decay-at-rest ($\pi$DAR) source used by the \textsc{Coherent} collaboration~\cite{COHERENT:2017ipa, COHERENT:2020iec, COHERENT:2024axu}, subsequently with solar neutrinos detected in the large-scale dark matter direct detection (DMDD) experiments \textsc{Xenon}nT~\cite{XENON:2024ijk}, \textsc{PandaX-4T}~\cite{PandaX:2024muv} and \textsc{LZ}~\cite{LZ:2025igz}, and finally with antineutrinos originating from a nuclear reactor by the \textsc{Conus} Collaboration~\cite{Ackermann:2025obx}.

The potential of using kg-size detectors for neutrino detection stimulated plenty of experimental activity with the goal of probing CE$\nu$NS with many different detection technologies and materials ~\cite{CONNIE:2021ggh, MINER:2016igy, NUCLEUS:2019igx, nGeN:2025hsd, NEON:2022hbk, RED-100:2024izi, RELICS:2024opj, Ricochet:2021rjo, Yang:2024exl, TEXONO:2024vfk}.

On theory side CE$\nu$NS has become an interesting tool for investigations within and beyond the standard model (BSM) because of its flavor-blind and, in principle, threshold-free properties~\cite{Papoulias:2017qdn,Cadeddu:2020lky,Miranda:2020tif,DeRomeri:2022twg,Alpizar-Venegas:2025wor,AtzoriCorona:2025ygn,Chattaraj:2025fvx}.
Within the SM it enables measurements of the Weinberg angle $\sin^{2}\theta_{W}$ at the MeV scale with neutrinos and probe modifications of the involved couplings, i.e.\ via radiative correction \cite{Canas:2018rng,AtzoriCorona:2023ktl,AtzoriCorona:2025xwr,AtzoriCorona:2026wbu} or investigation of the nuclear form factor when deviating from full coherence, i.e.\ with higher neutrino energies from $\pi$DAR sources~\cite{Amanik:2009zz,Patton:2012jr,AristizabalSierra:2019zmy,Coloma:2020nhf,Rossi:2023brv}.
BSM searches can be performed by testing for new neutrino interactions, for example in the context of heavy new physics via non-standard neutrino interactions (NSIs)\cite{Barranco:2005yy,Lindner:2016wff,Han:2020pff,Chatterjee:2022mmu,Denton:2022nol,Coloma:2023ixt,Lozano:2025ekx} or new light mediators~\cite{deNiverville:2015mwa,Dent:2016wcr,Bauer:2018onh,Farzan:2018gtr,Dent:2019ueq,Miranda:2020zji,AristizabalSierra:2020rom,Cadeddu:2020nbr,Demirci:2023tui,DeRomeri:2026prc}. 
Neutrino (electromagnetic) properties~\cite{Kosmas:2015sqa,Cadeddu:2018dux,Miranda:2019wdy,AtzoriCorona:2025ygn,Chattaraj:2025fvx} or emerging new particles may be probed as well~\cite{Brdar:2018qqj,Chang:2020jwl,Candela:2023rvt}.
Furthermore, future applications in the context of multi-messenger astronomy~\cite{Brdar:2018zds,Raj:2019wpy,Munoz:2021sad} or nuclear safeguarding seem promising~\cite{Bowen:2020unj,vonRaesfeld:2021gxl}. 
Experimental upscaling in the near future will allow such investigations via precision CE$\nu$NS measurements. 

As one of the leading Germanium-based experiments at a reactor site, \textsc{Conus} continuously improved its experimental reach and reported the first CE$\nu$NS detection with electron antineutrinos in 2025 \cite{Ackermann:2025obx}.
In first activities at the Brokdorf nuclear power plant in Germany (Kernkraftwerk Brokdorf, KBR) during the years 2018-2022, the strongest constraints on the SM CE$\nu$NS with a limit a factor of $1.6\sigma$ above expectation~\cite{CONUS:2020skt,CONUSCollaboration:2024kvo}, and limits on broad classes of BSM physics (NSIs, light mediators, neutrino electromagnetic properties) were obtained~\cite{CONUS:2021dwh,CONUS:2022qbb}. 
After the shutdown of the Brokdorf reactor and subsequent refining of the experimental configuration, the experiment transitioned to the nuclear power plant in Leibstadt, Switzerland (Kernkraftwerk Leibstadt, KKL)~\cite{CONUS:2024lnu}. 
With data collected there since 2023, a successful CE$\nu$NS detection was reported with $3.7\sigma$ significance. 

The present paper summarizes investigations of experimental data taken at both the Brokdorf and the Leibstadt site.
For the former, this includes searches for a neutrino magnetic moment ($\nu$MM) and neutrino millicharge ($\nu$MC) in an extended dataset (RUNs 1-5) as well as the latest dataset with an upgraded data acquisition system (DAQ) capable of pulse shape discrimination (PSD)~\cite{Bonet:2023kob}.
The first dataset of the latter - used to claim the first CE$\nu$NS detection at a reactor site - is now analyzed for further BSM physics~\cite{Ackermann:2025obx}.

This work is structured as follows: In section \ref{sec:exp_setup} the CONUS setup at the Brokdorf nuclear power plant and the CONUS+ setup at the Leibstadt nuclear power plant are introduced with special emphasis on the upgrades performed in the course of the experiment and during the transition to the new experimental site in Switzerland.
After that the BSM models investigated throughout this work are covered in section \ref{sec:bsm_topics} with corresponding results presented and discussed in section \ref{sec:results}. 
Finally, we conclude in section \ref{sec:conclusions} and give an outlook of the next experimental milestones.


\section{Experimental data collection and analysis framework}
\label{sec:exp_setup}

In this section the two experimental sites of the CONUS experiment and the applied analysis framework are introduced. 
We discuss the characteristics of the detector setups, describe the different datasets analyzed in this work and present the performed statistical investigations. 
Special emphasis is given to changes in the experimental configuration related to the transition from the Brokdorf (KBR) to the Leibstadt (KKL) nuclear power plant, i.e.\ the modifications that allowed for the first detection of CE$\nu$NS there.


\subsection{Experimental setup}

\paragraph{The Brokdorf site}

The \textsc{Conus} experiment was first set up at the commercial nuclear power plant in Brokdorf that housed a pressurized water reactor with a thermal power of $3.9$\,GW. 
Four p-type point contact high-purity germanium detectors~\cite{Bonet:2020ntx} with a total active mass of $(3.73\pm 0.02)$kg were deployed at a distance of 17.1\,m from the reactor's center.
At the experimental site the effective overburden was determined to be $\sim24$\,m of water equivalent (m w.e.).
Two different data acquisition systems were used for data collection. 
The integrated multichannel analyzer \textsc{Lynx} was used for the Germanium data taking in all experimental measurement campaigns, while it was supplemented by the digital multichannel analyzer module V1782 by \textsc{CAEN} in the last period of data collection. 
The latter allowed to lower the experimental threshold down to 210\,eV (previously $\sim300$\,eV) and to use pulse shape information for further background reduction.
Further details of the \textsc{Conus} detectors can be found in ref.~\cite{Bonet:2020ntx}.
The detectors were enclosed in a layered shield with passive and active background reduction components. In particular, lead, borated and non-borated polyethylene sheets were chosen to reduce the flux of gamma rays and neutrons reaching the detectors, respectively. 
Furthermore, plates of plastic scintillators equipped with photomultiplier tubes (PMTs) served as a muon veto anti-coincidence system.
Together with a flushing system, which is based on air low in airborne Rn content, an overall background reduction of a factor $10^{4}$ was achieved.
A complete description of the shielding capabilities and the background composition is given in \cite{Hakenmuller:2019ecb,Bonet:2021wjw}.

At the experimental location (17.1\,m distance from the reactor core) a neutrino flux of $2.3\cdot10^{13}\rm{cm}^{-2}\rm{s}^{-1}$ is expected with neutrino energies reaching up to $\sim10$\,MeV.
For the CE$\nu$NS signal prediction, a data-driven approach is chosen: the antineutrino emission is determined with measured antineutrino spectra and data provided by the operating company PreussenElektra GmbH, namely reactor thermal power and nuclear fuel composition.
We use antineutrino spectra provided by the Daya Bay collaboration~\cite{DayaBay:2021dqj} and add them according to the average fission fraction of the four main isotopes $(^{235}\text{U},\, ^{238}\text{U},\, ^{239}\text{Pu},\, ^{241}\text{Pu}) = (49.1,\, 7.4,\, 36.1,\, 7.4)\, \%$.
For the high energy part above 8\,MeV, a Daya Bay measurement of the summed spectrum is used~\cite{DayaBay:2022eyy}, while the low energy part below the threshold of inverse beta decay relies on knowledge gained from ab-initio calculations~\cite{Estienne:2019ujo}, with proper normalization of the whole spectrum taken into account.

\vspace{1cm}
\paragraph{The Leibstadt site} After the shutdown of the Brokdorf nuclear plant and subsequent background measurements, the whole experimental setup underwent refurbishment and was relocated to the Leibstadt nuclear power plant as new experimental site in 2023.
In this course, the detectors' detection threshold could be further lowered to energies of $160-180$\,eV. 
Using solely the \textsc{CAEN} data acquisition module for measurement, trigger efficiencies of almost 100\,\%  down to the individual detection thresholds are achieved.
The new experimental site inside the reactor building was selected and assessed with dedicated background measurement campaigns~\cite{CONUS:2024vyx}.
Due to a smaller amount of overburden, i.e.\ 7.4\,m w.e.\ compared to 24\,m w.e.\ in Brokdorf, the encountered cosmogenic background levels are higher than in the KBR analyses.
Furthermore, changes in background levels originating from an overburden variation of 0.3\,m w.e.\  are observed when the drywell head of the containment structure is placed above the experimental room (during reactor OFF times).
Improvements in the detector cooling system helped to reduce rate correlations with temperature (through microphonic noise).
Taking into account all the effects and adjusting the background models to the new experimental site, the background expectation both for reactor ON and OFF agree well with data~\cite{CONUS:2024vyx,Ackermann:2025obx}.

The experimental site is located 20.7\,m away from the core of the 3.6\,GW boiling water reactor and provides an antineutrino flux of $1.5\cdot10^{13}\rm{cm}^{-2}\rm{s}^{-1}$. 
Although the expected flux is $\sim 35\,\%$ lower than at the Brokdorf site, the overall improvement of detector components was assessed to allow for a successful CE$\nu$NS detection. 
Again, reactor thermal power evolution and fission composition are available for the analyses. 
For the first data collection, the average fuel compositions of the relevant emitting isotope chains were $(^{235}\text{U},\, ^{238}\text{U},\, ^{239}\text{Pu},\, ^{241}\text{Pu}) = (53,\, 8,\, 32,\, 7)\, \%$.
Further details of the new experimental site can be found in refs.~\cite{CONUS:2024vyx,CONUS:2024lnu}. The main characteristics of both reactor sites are summarized in table \ref{tab:site_properties}.


\begin{table}[t]
    \centering
    \begin{tabular}{c c c}
        \hline
        & KBR & KKL \\
        \hline
        Thermal Power & $3.9 \ \mathrm{GW}$ & $3.6 \ \mathrm{GW}$ \\
        Antineutrino flux & $2.3 \cdot 10^{13} \ \mathrm{cm^{-2} s^{-1}}$ & $1.5 \cdot 10^{13} \ \mathrm{cm^{-2} s^{-1}}$ \\
        Distance from core & $17.1 \ \mathrm{m}$ & $20.7 \ \mathrm{m}$ \\
        Overburden & $24 \ \mathrm{m\ w.e.}$ & $7.4 \ \mathrm{m\ w.e.}$ \\
        \hline
    \end{tabular}
    \caption{Main experimental properties of the two reactor sites in Brokdorf and Leibstadt.}
    \label{tab:site_properties}
\end{table}


\subsection{Latest datasets of the CONUS experiments}\label{sec:datasets}

In this work, three different datasets are analyzed: an extended dataset with the \textsc{Lynx} DAQ (\textit{KBR Lynx}) investigated in the context of neutrino electromagnetic properties, the last dataset collected at the KBR site assembled with the new \textsc{CAEN} DAQ (\textit{KBR CAEN}) and the first dataset of the new KKL site (\textit{KKL CAEN}).
Certain criteria on the detection threshold and postprocessing cuts were applied to ensure appropriate stability and quality of the obtained datasets. 
In doing so, noise events triggered by room temperature fluctuations and spurious DAQ-induced events were removed.
Conservative assumptions on the detectors' threshold as listed in table \ref{tab:energy_regions} prevent the leakage of detector noise into the region of interest (ROI) used in the analyses.
Further details of the data processing can be found in refs.~\cite{CONUS:2024lnu, Ackermann:2025obx}. 

\vspace{0.1cm}
\noindent In the following, characteristics of the individual datasets are introduced:


\begin{table}[]
    \centering
    \begin{tabular}{c c c}
    \hline
    Energy region &  Energy & Models\\
    \hline
    Low & $E_{th}-800 \ \rm{eV}$ & All \\
    Mid & $2-8 \ \rm{keV}$ & NMM / Light mediators \\
    High & $12-20 \ \rm{keV}$ & NMM (Lynx)\\
    \hline
    \end{tabular}
    \caption{Definition of the energy regions used in the respective BSM analyses. $E_{th}$ individually denotes the threshold of the detectors.}
    \label{tab:energy_regions}
\end{table}


\paragraph{\textit{KBR Lynx}}

The measurement campaign of the \textsc{Lynx} DAQ covers five data collection periods (runs) at KBR. For this dataset, data of Run-1 and Run-2 (2018 to 2019) are combined with data of Run-4 and Run-5 (2020 to 2022). 
The Run-1/Run-2 data already analyzed in \cite{CONUS:2021dwh} exhibit an exposure of $689.1\,\mathrm{kg\cdot d}$ reactor ON and $131.0\,\mathrm{kg\cdot d}$ reactor OFF data, covering a ROI of $2-8\,\mathrm{keV}$. 
After a pressure test of the reactor containment in June 2019, a new continuous background contribution below 10\,keV appeared leading two background levels up to a factor of 2 higher than before. 
Despite several tests and simulations, the source of this additional component remained unknown. 
Consequently, it was modeled as additional component with an analytic function of three parameters.
Data of Run-3 are excluded in this analysis as they were used for experimental optimization.
For the Run-5 data, an extended OFF-time of roughly one year is available due to the reactor's shutdown much higher that the standard yearly outage period of one month. 
The obtained exposure is $897.59\,\mathrm{kg\cdot d}$ reactor ON and $693.57\,\mathrm{kg\cdot d}$ reactor OFF data with an extended ROI of $2-20\,\mathrm{keV}$. 
Run-5 exhibits the largest exposure of reactor ON and OFF periods and is the most stable KBR run in terms of environmental and detector conditions due to an improved setup. 
In this run, no data of the detector CONUS-3 is used, as it was utilized to monitor the radon content inside the shield.
This dataset is used to get constraints on the neutrino magnetic moment and millicharge.


\paragraph{\textit{KBR CAEN}}

This dataset is obtained from the last data collection period (Run-5) at the Brokdorf site which started in May 2021 after major experimental refinements.
These include the installation of the \textsc{CAEN} DAQ and measures to further stabilize the experimental condition on site.
The data contain an extended background measurement after the reactor shutdown by the end of 2021, covering an effective exposure of 426\,$\si{kg}\cdot\si{d}$ reactor on and 272\,$\si{kg}\cdot\si{d}$ reactor off (in detector active mass).
As described for the \textit{KBR Lynx} dataset, one of the detectors was excluded from the analysis as it was used to monitor the radon level inside the shield.
With the \textsc{CAEN} DAQ, detector thresholds could be lowered from 300\,eV down to 210\,eV and pulse shape information of the recorded signals allowed for further background discrimination in the ROI~\cite{Bonet:2023kob}.
Knowledge from data taken with the \textsc{Lynx} DAQ allows to further constrain the normalization of the applied background model to an uncertainty of 10\%.
As previously mentioned for the \textit{KBR Lynx} dataset, after a pressure test of the reactor containment, a new background contribution appeared and was, here, modeled with an analytic function of two parameters.
This dataset imposed an overall upper limit on the CE$\nu$NS of $<143$ antineutrinos (at 90\,\% C.L.), corresponding to $<0.34\,\rm{kg}^{-1}\rm{d}^{-1}$ and only a factor of 2 above the SM expectation, and improved previous results by more than one order of magnitude~\cite{CONUS:2020skt}.
Furthermore, it allowed to constrain the amount of signal quenching in the Germanium, i.e.\ exclude certain $k$ parameters of the underlying model $k>0.21$ (at 90\,\% C.L.), where $k$ roughly corresponds to the ratio of ionization to nuclear recoil energy at 1\,keV nuclear recoil energy.
Below we will investigate this dataset for its capabilities of constraining vector NSI and light mediator particles.
For the latter, both the \textit{low} and \textit{mid} energy regions will be used simultaneously in the analysis. cf.~table~\ref{tab:energy_regions}.



\begin{figure}[t]
    \centering
    \includegraphics[trim=0 12 0 0,clip, width=0.8\linewidth]{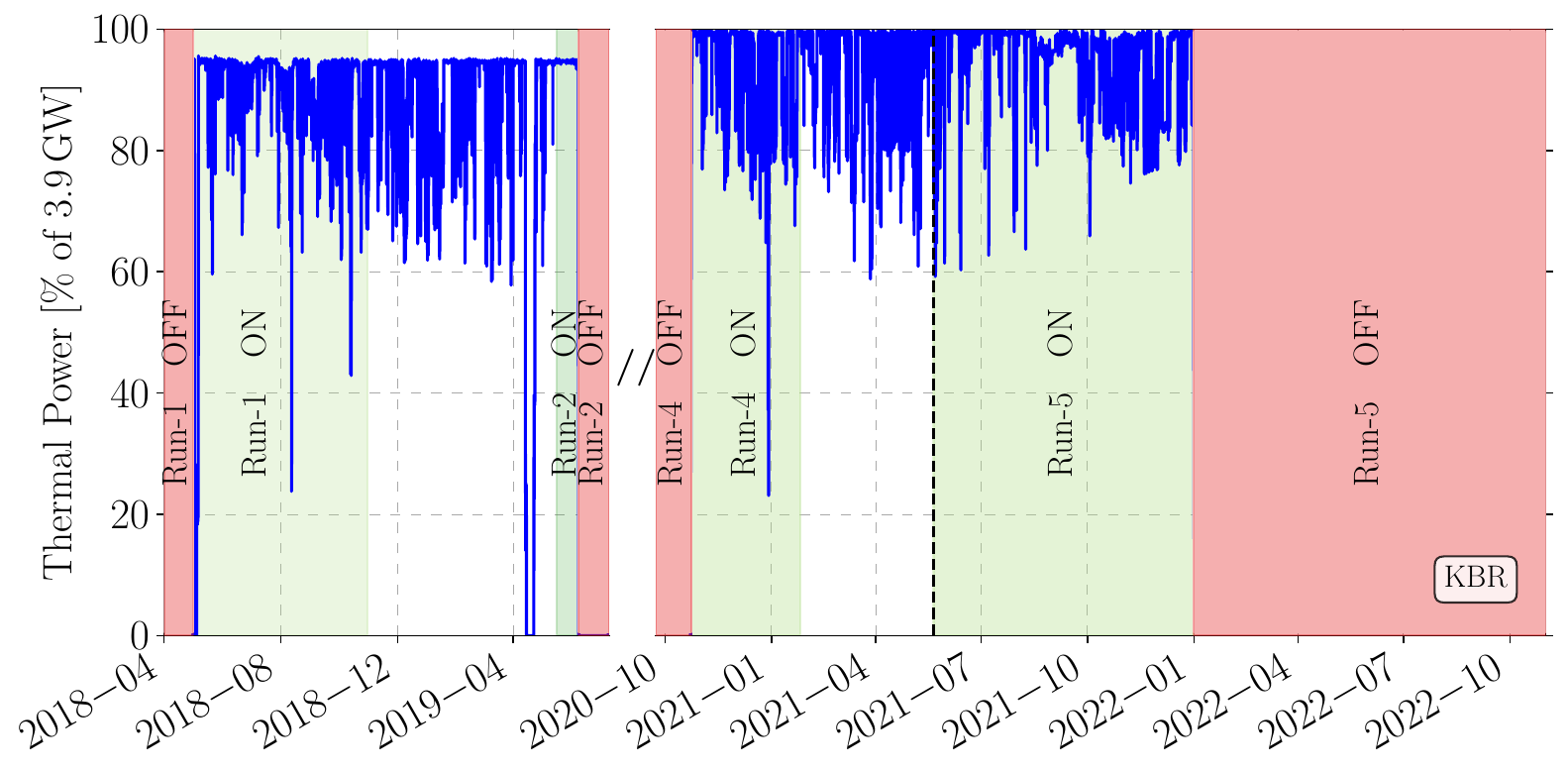}
    \includegraphics[width=0.8\linewidth]{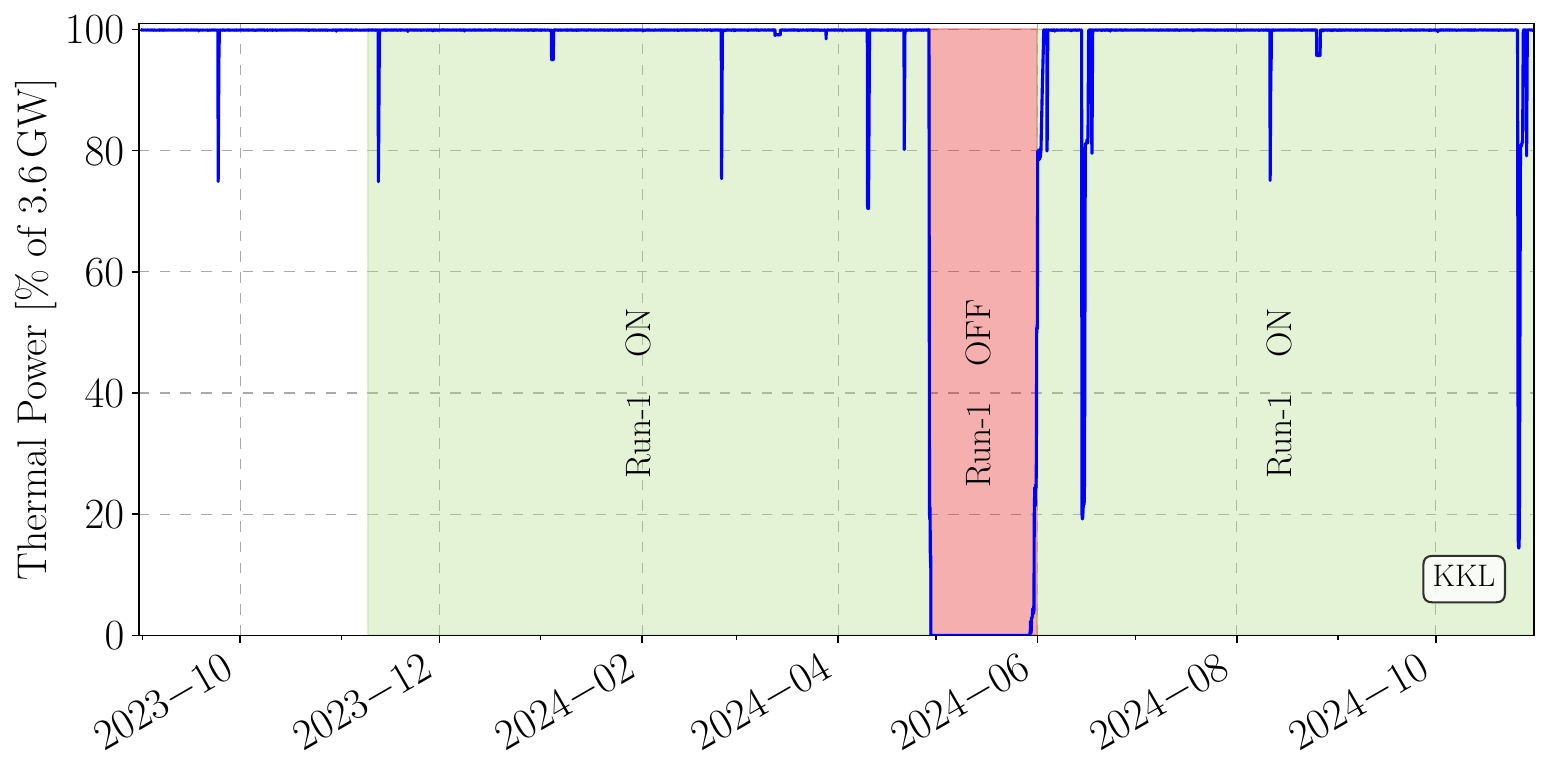}
    \caption{Data collection campaigns (including reactor thermal powers in blue) at the nuclear power plants in Brokdorf (top) and Leibstadt (bottom). The period June 2019-October 2020 was used for experimental optimization and is not used in the analyses.}
    \label{fig:data_collection}
\end{figure}


\paragraph{\textit{KKL CAEN}}

After commissioning at the KKL site, the first data were taken from November 2023 to July 2024 with a reactor outage in May 2024. 
With three detectors - one excluded due to stability issues - an overall detector mass of $(2.83\pm0.02)$\,kg was available. 
The CE$\nu$NS ROI has been adjusted from the detector threshold up to 800\,eV, while high energy data analyzed in this work were taken up to 8\,keV.
Although available with the \textsc{CAEN} DAQ, PSD has not been applied for this data collection period. 
Nevertheless, the data collected with the refined experimental setup allowed to measure ($395\pm106$) CE$\nu$NS events with a significance of $3.7\sigma$ and satisfying agreement with the SM prediction.
As for the \textit{KBR CAEN} data, this dataset will be used to constrain vector NSI, light mediators and electromagnetic properties (in conjunction with the \textit{mid} ROI dataset).

\vspace{0.25cm}
\noindent Illustrations of the data collection periods both at KBR and KKL, and details of the individual datasets and detectors are shown in figure~\ref{fig:data_collection} and table~\ref{tab:exposures_cenns}.


\begin{table}[t]
    \centering
    \resizebox{\linewidth}{!}{
    \begin{tabular}{c c c c c}
    \hline
    Datasets  & Energy region & Detector   &  reactor ON ($\si{kg} \cdot \si{d}$) & reactor OFF ($\si{kg} \cdot \si{d}$) \\
    \hline
    \textit{KBR Lynx}   & Mid & \textit{C}1 (Run-1)  &  215.4 &  29.6 \\
                        &   & \textit{C}2 (Run-1)  &  184.6  &  32.2 \\
                        &   & \textit{C}3 (Run-1)  &  248.5  &  31.7 \\
                        &   & \textit{C}1 (Run-2)  &  19.8  &  18.5 \\
                        &   & \textit{C}3 (Run-2)  &  20.8  &  19.0 
                        \vspace{2mm}\\
                        && Total  & 689.1  & 131.0 \\
                        \cline{2-5}
                        & Mid/High & \textit{C}1 (Run-4)  &  69.46  &  21.15 \\
                        &   & \textit{C}2 (Run-4) &  56.92  &  21.86 \\
                        &   & \textit{C}3 (Run-4) &  69.32  &  21.30 \\
                        &   & \textit{C}4 (Run-4) &  69.42  &  27.53 \\
                        &   & \textit{C}1 (Run-5) &  215.26  &  286.10 \\
                        &   & \textit{C}2 (Run-5) &  214.60  &  43.77 \\
                        &   & \textit{C}4 (Run-5) &  202.61  &  271.86 
                        \vspace{2mm} \\
                        && Total  & 897.59  & 693.57 \\
    \hline
    \textit{KBR CAEN}   & Low/Mid & \textit{C}1 ($E_{th} = 210 \ \rm{eV}$) &  141.5  &  40.2 \\
                        &   & \textit{C}2 ($E_{th} = 210 \ \rm{eV}$) &  145.5  &  130.3 \\
                        &   & \textit{C}4 ($E_{th} = 210 \ \rm{eV}$) &  139.0  &  101.6 
                        \vspace{2mm} \\
                        && Total  & 426.0  & 272.1 \\
    \hline\hline
    \textit{KKL CAEN}   & Low/Mid & \textit{C}2 ($E_{th} = 180 \ \rm{eV}$)  &  111.2  &  19.8 \\
                        &   & \textit{C}3 ($E_{th} = 160 \ \rm{eV}$) &  103.3  &  19.9 \\
                        & & \textit{C}5 ($E_{th} = 170 \ \rm{eV}$) &  112.3  &  20.0 
                        \vspace{2mm} \\
                        && Total  & 326.8  & 59.7 \\
    \hline
    \end{tabular}
    }
    \caption{Detailed exposures and regions of interest for the individual detectors and the three datasets under study (quality cuts included). The energy regions are defined in table~\ref{tab:energy_regions}.}
    \label{tab:exposures_cenns}
\end{table}



\subsection{Standard model reaction channels and analysis framework}\label{analysis_framework}

Reactor antineutrinos can undergo two SM interactions in the detector material: coherent elastic neutrino-nucleus scattering (CE$\nu$NS) and elastic neutrino-electron scattering (E$\nu$eS). 
While measuring the first is the experiment's prime objective, the latter remains basically unobserved due to the small target mass and the neutrino's genuine weakly interacting nature within the SM.
However, it can be used to constrain sizable couplings in the BSM context, as we will show below.

In CE$\nu$NS, the neutrino scatters off a whole nucleus, which happens at low momentum transfer when the wavelength of the mediating particle, i.e.\ a $Z$ boson, is of similar size than the target nucleus. 
In this case, the individual scattering amplitudes add up coherently and allow for an enhanced interaction. 
The corresponding interaction cross section is given by
\begin{equation}\label{eq:cross_section_sm_cenns}
\begin{aligned}
    \frac{d\sigma}{dT_{A}}(T_{A}, E_{\nu}) & = \frac{G_{F}^{2} m_{A}}{4 \pi} Q_{W}^{2} \left(1-
\frac{m_{A} T_{A}}{2 E^{2}_{\nu}}\right) |F(T_{A})|^{2}\, , \\ 
\text{with } Q_{W} &=\left[(1-4\sin^{2}\theta_{W})Z - N\right]\, ,
\end{aligned}
\end{equation}
with the neutrino energy $E_{\nu}$, the recoil energy $T_{A}$ and mass $m_{A}$ of the struck nucleus. 
The weak nuclear charge $Q_{W}$ exhibits a mild dependence on the Weinberg angle $\sin^{2}\theta_{W}$ and depends on the number of protons $Z$ and neutrons $N$, respectively. 
Hence, the cross section mainly scales with the squared number of neutron of the scattering target.
The nuclear form factor $F(T_{A})$ describes the deviation from scattering off a point-like target and is used here in the Helm parameterization~\cite{Engel:1991wq}
\begin{align}
    F(T_{A}) = \frac{3j_{1}(q(T_{A}) R_{1})}{q(T_{A}) R_{1}}\exp\left[-\frac{(q(T_{A})s)^{2}}{2}\right]\, ,
\end{align}
with the spherical Bessel function $j_{1}$, the momentum transfer depending on target mass and nuclear recoil energy $q_{2} = 2m_{A}T_{A}$, the nuclear skin thickness $s\simeq1$\,fm and the radius parameters $R_{1}=\sqrt{R^{2} - 5s^{2}}$ and $R\simeq 1.2\sqrt[3]{A}$\,fm.

The electroweak cross section for E$\nu e$S mediated by $W$ and $Z$ bosons is given by~\cite{Vogel:1989iv}
\begin{align}
    \frac{d \sigma}{d T_e} (T_e, E_\nu) = \frac{G_F^2 m_e}{\pi}  \left[ (g_V + g_A)^2 + (g_V-g_A)^2 \left( 1 - \frac{T_e}{E_\nu}\right) - (g_A^2 -g_V^2) \frac{m_e T_e}{2 E_\nu^2}\right]\, ,
    \label{eq:enes_cross}
\end{align}
with the vector and axial-vector couplings for antineutrinos scattering of electrons $g_V=\frac{1}{2} + 2 \sin^2{\theta_W}$ and $g_A = -\frac{1}{2}$, respectively.

The maximal recoil energy for both reactions is given by
\begin{align}
    T_{\rm{max}} = \frac{2 E_{\nu}^2}{m_{x}+2 E_{\nu}}\, ,
    \label{eq:Tmax}
\end{align}
with the antineutrino energy $E_{\nu}$ and the target mass $m_{x}$ with $x={\{e, A\}}$.
For a given recoil (energy) this expression determines which (anti-)neutrino energies contribute in the interactions.\footnote{Technically, it defines the lower neutrino energy used when weighting the cross sections above with the reactor antineutrino spectrum: $\frac{d\bar{\sigma}}{dT_{x}}(T_{x}) = \int_{E_{\rm{min}(T_{x})}}^{E_{\rm max}} dE_{\nu} \frac{dN}{E_{\nu}}(E_{\nu})\frac{d\sigma}{dT_{x}}(T_{x},E_{\nu})$.}

The germanium detectors of the \textsc{Conus} setup measure ionization signals, so the nuclear recoils created in CE$\nu$NS are not directly assessable. 
Only $\lesssim 20$\% of the initial energy can be recorded, while the remaining energy is lost in dissipative processes. 
This so-called signal quenching in germanium can be described by the Lindhard model~\cite{Lindhard:1961zz}.
For this analysis, we use measurements of the model's parameter $k$ presented in \cite{Bonhomme:2022lcz}, i.e.\ $k=0.162\pm 0.004$ where $k$ describes the ratio between obtained ionization and recoil energy at roughly 1\,keV recoil energy. 
Good agreement between the model and data have been confirmed by an independent measurement~\cite{Li:2022xbv}.
Furthermore, the binding energies of electrons need to be considered, thus, we use the same binding energy correction already applied in previous analyses~\cite{CONUS:2021dwh}.
To account for the detector response, detection efficiencies and resolutions are individually applied for each detector.  
In the case of \textit{KBR CAEN}, the effect of PSD is included as well~\cite{Bonet:2023kob}.

The background models consist of several components: a Monte Carlo simulation of the intrinsic background level expected at the experimental rooms, the additional component that appeared after the mentioned  pressure test at KBR, and an additional component that account for inefficiencies in the muon veto system at KBR. 


\begin{table}[t]
    \centering
    \resizebox{\linewidth}{!}{
    \begin{tabular}{c c|c}
    \hline
      & Quantity & Uncertainty or related parameter \\
      \hline
      \textit{KBR Lynx} & background MC & $\Theta_{b}$ ($\leq 10\,\%$, uncertainty from background model) \\
      & reduced neutrino flux $\Delta \Phi^{*}$ & $\sim \! 3\,\%$ \\
      & neutrino spectrum & subdominant uncertainty\\
      & reactor \textsc{On} and \textsc{Off} duration & negligible uncertainty \\
      & active mass & $< 1 \,\%$ \\
      & electronic detection efficiency $c_\mathrm{eff}$ &  $\leq5\%$ \\
      & energy calibration uncertainty $\Delta E$ & 15\,eV$_{ee}$ \\
      \hline
      \textit{KBR CAEN} & background MC & $\Theta_{b}$ ($\leq 10\,\%$, uncertainty from background model) \\
      & reduced neutrino flux $\Delta \Phi^{*}$ & $\sim \! 3\,\%$ \\
      & neutrino spectrum & subdominant uncertainty (compared to quenching)\\
      & reactor \textsc{On} and \textsc{Off} duration & negligible uncertainty \\
      & active mass & $< 1 \,\%$ \\
      & trigger efficiency & $\mu_{\rm eff}, \sigma_{\rm eff}$, $<8\%$ \\
      & PSD efficiency & $\mu_{\rm PSD}, \sigma_{\rm PSD}$, $<3\%$ \\
      & PSD background reduction & $\sim 6\,\%$ \\
      & energy calibration uncertainty $\Delta E$ & 5\,eV$_{ee}$ \\
      & quenching & $k\sim 2 \,\%$\\
      \hline \hline
      \textit{KKL CAEN} & background MC & $\Theta_{b}$ ($\leq 10\,\%$, uncertainty from background model) \\
      & reduced neutrino flux $\Delta \Phi^{*}$ & $\sim \! 2.5\,\%$ \\
      & neutrino spectrum & subdominant uncertainty (compared to quenching)\\
      & reactor \textsc{On} and \textsc{Off} duration & negligible uncertainty \\
      & active mass & $< 1 \,\%$ \\
      & trigger efficiency & $\mu_{\rm eff}, \sigma_{\rm eff}$, $<0.9\,\%$ \\
      & energy calibration uncertainty $\Delta E$ & 5\,eV$_{ee}$ \\
      & quenching & $k\sim 2\,\%$
    \end{tabular}
    }
    \caption{ Overview of the quantities entering the likelihood function with corresponding uncertainties.}
    \label{tab:likelihood_uncertainty}
\end{table}


In the statistical investigation, reactor ON and OFF data of the detectors considered are simultaneously fitted in a maximum likelihood approach with model parameters shared among all likelihood functions. 
Correlated and uncorrelated experimental uncertainties are incorporated with (Gaussian) pull terms.
The overall likelihood function (for one detector) takes the form 
\begin{equation}\label{eq:likelihood_full}
    \begin{aligned}
        -2\log\, \mathcal{L} &= -2\log\, \mathcal{L}_{\rm ON} -2 \log\, \mathcal{L}_{\rm OFF} \\ 
    & + (\mathbf{\theta} - \mathbf{\bar{\theta}})^{T} \rm{Cov}^{-1} (\mathbf{\theta} - \mathbf{\bar{\theta}}) + 2 \sum_i \frac{(\theta_{i} - \theta_i^*)^2}{2 \sigma_{i}^2}\, ,
    \end{aligned}
\end{equation}
assuming Poissonian binned likelihood functions $\mathcal{L}_{\rm ON, OFF}$. 
External knowledge from auxiliary measurements is included via the pull terms in the second line. 
Cov represents the covariance matrix of correlated parameters (e.g.\ quantities describing trigger efficiency) and $\theta^{*}_{i}$ and $\sigma_{i}$ the mean and standard deviation of uncorrelated parameters (e.g.\ the $k$ parameter of the Lindhard model).
An overview of all fit parameters with uncertainties for the individual datasets can be found in table~\ref{tab:likelihood_uncertainty}.

We perform a likelihood ratio tests~\cite{Cowan:2010js} for the parameters of the models under study and determine limits at 90\% C.L.\ from a $\chi^{2}$-distributed test statistics.
In contrast to previous (BSM) results, information about signal quenching with corresponding uncertainty is directly incorporated.
Besides this, the analysis framework is based on previously published SM CE$\nu$NS results~\cite{CONUS:2020skt,CONUS:2021dwh,CONUS:2022qbb,CONUSCollaboration:2024kvo,Ackermann:2025obx}.


\section{New physics beyond the Standard Model in \texorpdfstring{CE$\nu$NS}{CEνNS} experiments}
\label{sec:bsm_topics}

In the following, we introduce the different phenomenological topics that can be studied with the available experimental data. 
We discuss extensions of the standard model such as non-standard interactions and new light mediators, but also cover electromagnetic properties of the neutrinos and the determination of the Weinberg angle $\sin^{2}\theta_{W}$.
Especially, deviations of the latter from its SM expectation may be a smoking gun signature for new physics at work.
The ordering of topics follows the analyses performed with the different datasets.


\subsection{Electromagnetic neutrino properties}

With oscillation experiments confirming the massive nature of neutrinos, new properties may emerge from SM extensions, such as electromagnetic properties. 
Among them, a non-zero neutrino magnetic moment (NMM) and a potential millicharge (NMC) have attracted special attention. 
In particular, a detection of a neutrino magnetic moment could help to discriminate between Majorana and Dirac neutrinos~\cite{Bell:2006wi,Giunti:2014ixa}.
Further, finding a larger value than expected from minimal extension (SM with three right-handed neutrinos) would hint to more complex new physics. 
The signature of these properties contribute additional events with different spectral behavior.
For the neutrino magnetic moment $\mu_\nu$ these additional contributions follow \cite{Giunti:2014ixa,Papoulias_2018}:
\begin{align}\label{eq:nmm_cross_section}
    &\left( \frac{d \sigma}{d T_A} \right)_{\text{CE$\nu$NS} +\text{NMM}} = \left(\frac{d \sigma}{dT_A}\right)_{\text{CE$\nu$NS}} + \pi Z^{2}\frac{\alpha^2 \mu_{\nu}^{2}}{m_{e}^{2}} \left( \frac{1}{T_A} - \frac{1}{E_\nu}\right)\, ,\\
    &\left(\frac{d\sigma}{dT_{e}}\right)_{\text{E$\nu$eS} +\mathrm{NMM}} = \left(\frac{d \sigma}{dT_e}\right)_{\text{E$\nu$eS}}+ \pi \frac{\alpha^{2}}{m_{e}^{2}} \left(\frac{1}{T_{e}} - \frac{1}{E_{\nu}}\right) \left(\frac{\mu_{\nu}}{\mu_{B}}\right)^{2}\, ,
\end{align}
where $\alpha$ is the fine structure constant, $m_e$ the electron mass and $\mu_B$ Bohr's magneton.
Scatterings induced by a non-zero NMC will lead to the following modifications~\cite{Brudanin:2014iya,Cadeddu:2020lky}
{\small
\begin{align}\label{eq:nmc_cross_section}
    &\left(\frac{d \sigma}{dT_A}\right)_{\text{CE$\nu$NS} + \text{NMC}} = \frac{G_F^2 M}{\pi} \left( 1 - \frac{M T_A}{2E_\nu^2} \right) \left[\left(g_V^p-\frac{\pi \alpha}{\sqrt{2} G_F M^2 T_A^2}\right) Z + g_V^n N \right]^2 |F(T_A)|^2\, , \\
    &\left(\frac{d \sigma}{dT_A}\right)_{\text{E$\nu$eS} + \text{NMC}} \approx \left(\frac{d \sigma}{dT_A}\right)_{\text{E$\nu$eS}} + \frac{2 \pi \alpha}{m_e T^2} q_\nu^2 \, .
\end{align}
}
Requiring the E$\nu$eS cross section induced by a NMC to be smaller than the one for a NMM already allows to estimate a limit on the NMC from a given NMM~\cite{Brudanin:2014iya}:
\begin{align}\label{eq:nmc_limit_from_nmm}
    q_{\nu}^2<\frac{T}{2m_e}\left(\frac{\mu_{\nu}}{\mu_B}\right)^2 e_0 \, ,
\end{align}
where $e_0$ is the elementary charge. 
This approximation has been used before \cite{Brudanin:2014iya,CONUS:2022qbb} and will be used for the \textit{KBR Lynx} dataset. 

Finally, a sizable neutrino charge radius would modify existing couplings and, thus, shift the Weinberg angle away from its SM expectation~\cite{Papoulias_2018,Giunti:2014ixa}
\begin{align}
    \sin^2{\theta_W} \rightarrow \sin^2{\bar{\theta}_W} + \frac{\sqrt{2} \pi \alpha}{3 G_F} \langle r_{\nu_e}^2\rangle \, ,
    \label{eq:charge_radius}
\end{align}
where $\bar{\theta}_W$ is the Standard Model value.
This underlines the importance of determining the Weinberg angle with (anti-)neutrinos.


\subsection{Non-standard interactions of vector-type}

A model-independent way to look for new neutrino interactions originating from ultraviolet (UV, i.e. BSM effects of high energy origin) physics is offered by the framework of non-standard interactions in the neutrino-quark sector. 
Therein, heavy new particles are integrated out such that new interactions are conventionally described in terms of 4-Fermi interactions. 
The corresponding vector-type operators that couple neutrinos and quarks generally take the form 
\begin{align}\label{eq:operator_vector_nsi}
    \mathcal{O}^{q \mathrm{V}}_{\rm NSI}=\left(\bar{\nu}_{\alpha} \gamma^{\mu} L \nu_{\beta}
 \right)\left(\bar{q} \gamma_{\mu}Pq  \right) + \text{h.c.}\, ,
\end{align}
where $P=\{{P_{L},P_{R}}\}$ are the left- and right-handed projection operators, $\alpha, \beta \in \{\text{e}, \mu, \tau\}$ and $\gamma_\mu$ the Dirac matrices.
If present, these new interactions interfere with the SM CE$\nu$NS interaction and, thus, modify the weak nuclear charge in Eq.~\eqref{eq:cross_section_sm_cenns} to
\begin{equation}
\begin{aligned}
	\mathcal{Q}_{\rm NSI}^{\mathrm{V}} =& \left( g_{V}^{p} + 2 \epsilon^{u\mathrm{V}}_{\alpha \alpha} 
	+ \epsilon^{d\mathrm{V}}_{\alpha \alpha} \right) Z + \left(g_{V}^{n} + \epsilon^{u\mathrm{V}}_{\alpha \alpha} + 2 \epsilon^{d\mathrm{V}}_{\alpha \alpha}\right) N \\
	&+ \sum_{\alpha, \beta} \left[ \left( 2\epsilon^{u\mathrm{V}}_{\alpha\beta} + \epsilon^{d\mathrm{V}}_{\alpha\beta} \right)Z + \left( \epsilon^{u\mathrm{V}}_{\alpha \beta } + 2 \epsilon^{d\mathrm{V}}_{\alpha \beta} \right)N  \right]\, ,
\end{aligned}
\label{eq:weak_nsi}
\end{equation}
with the vector coupling to neutrons $g_{V}^{n}=-\frac{1}{2}$ and protons $g_{V}^{n}=\frac{1}{2} -2\sin^{2}\theta_{W}$, respectively.
Depending on the values of the introduced couplings $\epsilon$, constructive and destructive interference between the new and the SM interaction is possible.
Consequently, the presence of NSIs can affect the CE$\nu$NS signal's overall normalization.
Limits on these new couplings can be directly translated into bounds on the energy scale $\Lambda_{\rm NP}$ where the new physics is expected to become dynamical: $\Lambda_{\rm NP} \sim m_{W}/\sqrt{\epsilon}$ with $m_{W}$ being the mass of the $W$ boson.  
Since nuclear reactors emit electron-antineutrinos, we only consider electron-type couplings and, for simplicity, ignore flavor transitions.
Hence, the parameter space is spanned by the parameter pair $( \epsilon^{u\mathrm{V}}_{ee},\ \epsilon^{d\mathrm{V}}_{ee})$.


\subsection{Light mediator models}

Next to heavy new physics, we further investigate the effect of light mediating particles coupled to neutrinos in the framework of so-called simplified models~\cite{Cerdeno:2016sfi}.
When light enough, these particles are dynamical and thus interactions explicitly depend on the momentum transfer they carry. 
This not only affects the normalization of the expected signal, but also allows for spectral distortions, which, when resolvable, allow for a discrimination with NSIs. 
In this context, we test modifications to E$\nu$eS as well and consequently add data collected at higher energies.


\subsubsection*{Light vector mediator}

One of the simplest and widely tested extensions of the SM is a gauged $U(1)$ interaction. 
Here, we assume the corresponding exchange boson to couple only to the first generation of SM fermions as described by the following Lagrangian
\begin{align}\label{eq:lagrangian_vector_mediator}
	\mathcal{L}_{Z'} = Z'_{\mu} \left( g^{\nu \mathrm{V}}_{Z'} \bar{\nu}_{L} \gamma^{\mu} \nu_{L}
	+  g^{e \mathrm{V}}_{Z'} \bar{e} \gamma^{\mu} e  + g^{q \mathrm{V}}_{Z'} \bar{q} \gamma^{\mu} q \right) 
	+ \frac{1}{2} m^{2}_{Z'} Z'_{\mu}Z'^{\mu}\, ,
\end{align}
where we stay agnostic about the generation of the vector boson mass $m_{Z'}$.
We focus on two benchmark cases widely studied in the CE$\nu$NS literature: coupling to $B-L$ charge and universal couplings among all fermions $g_{Z'} \equiv g^{\nu \mathrm{V}}_{Z'}=g^{e \mathrm{V}}_{Z'}=g^{u \mathrm{V}}_{Z'}=g^{d \mathrm{V}}_{Z'}$. 
The corresponding interactions interfere with the SM CE$\nu$NS cross sections and modify the cross section to
\begin{align}\label{eq:cross_section_vector_b_l}
    \frac{d\sigma}{dT_{A}}(T_{A}, E_{\nu}) & = \frac{G_{F}^{2}m_{A}}{\pi} \bigg[ Q_{W} - \frac{ g_{Z'}^{2}}{\sqrt{2}G_{F}} \frac{Z+N}{2m_{A}T_{A} + m_{Z'}^{2}}\bigg]^{2} \left(1-
\frac{m_{A} T_{A}}{2 E^{2}_{\nu}}\right) |F(T_{A})|^{2}\, ,
\end{align}
for the $U(1)_{B-L}$ extension of the SM and
\begin{align}\label{eq:cross_section_vector_universal}
    \frac{d\sigma}{dT_{A}}(T_{A}, E_{\nu}) & = \frac{G_{F}^{2}m_{A}}{\pi} \bigg[ Q_{W} + \frac{3 g_{Z'}^{2}}{\sqrt{2}G_{F}} \frac{Z+N}{2m_{A}T_{A} + m_{Z'}^{2}}\bigg]^{2} \left(1-
\frac{m_{A} T_{A}}{2 E^{2}_{\nu}}\right) |F(T_{A})|^{2}\, ,
\end{align}
for the universally coupled model. 
Notice that the sign difference for the latter allows both constructive and destructive interference (as $Q_{W}<0$), while the $B-L$ models only allows for a signal increase.
Furthermore, in the universally coupled case, there exists a band in parameter space which cannot be distinguished from the SM, i.e.\ $[Q_{W}+ {\rm BSM} ]\simeq -Q_W$.

For E$\nu$eS the modifications from both models are equal and given by
\begin{align}\label{eq:cross_section_ligth_vector_nu_e}
    \left(\frac{d\sigma}{dT_{e}}\right)_{{\rm E\nu eS} + Z'}=\left(\frac{d\sigma}{dT_{e}}\right)_{{\rm E\nu eS}} 
    + \frac{\sqrt{2} G_{F} m_{e} g_{V} g_{Z'}^{2} }{\pi (2m_{e}T_{e}  + m_{Z'}^{2})} 
    + \frac{m_{e}  g_{Z'}^{4} }{2\pi (2m_{e}T_{e}  + m_{Z'}^{2})^{2}}\, ,
\end{align}
with the antineutrino electron coupling $g_{V}$. 
With a good enough sensitivity to this interaction channel, there is the potential to lift the degeneracy present in nuclear scattering via a light universally coupled vector mediator when the relevant part of parameter space is cut by the E$\nu$eS sensitivity.


\subsubsection*{Light scalar bosons}

To test the modification of an interaction induced by a light scalar boson, we assume again only a coupling to the first generation as described by the following Lagrangian
\begin{align}\label{eq:lagrangian_light_scalar}
    \mathcal{L}_{\phi} = \phi\left( g^{q \mathrm{S}}_{\phi} \bar{q}q + g^{e \mathrm{S}}_{\phi} \bar{e}e + g^{\nu \mathrm{S}}_{\phi} \bar{\nu}_{R} \nu_{L}  + \text{h.c.} \right) 
    - \frac{1}{2} m^{2}_{\phi} \phi^2\, ,
\end{align}
with the boson mass $m_{\phi}$ and the individual scalar coupling $g^{x \mathrm{S}}_{\phi}$ for $x = \{\nu,\, e,\, q\}$ and $q = \{u,\, d\}$.
To further simplify the analysis, universal couplings will be assumed, thus setting $g_\phi^{qS} = g_\phi^{eS} = g_\phi^{\nu S} \equiv g_{\phi}$.
Note that the Yukawa interaction cannot interfere with the vector interaction of SM CE$\nu$NS due to different final states. 
The modified CE$\nu$NS cross section is the sum of both contributions:
\begin{align}\label{eq:cross_section_light_scalar_cenns}
    \left(\frac{d\sigma}{dT_{A}}\right)_{\rm CE\nu NS + \phi} = \left(\frac{d\sigma}{dT_{A}}\right)_{\rm CE\nu NS} 
    + \frac{(g^{\nu \mathrm{S}}_{\phi} \mathcal{Q}_{\phi})^{2} m_{A}^{2} T_{A} }{4\pi E_{\nu}^{2} (2 m_{A} T_{A}  + m_{\phi}^{2})^{2}}\, ,
\end{align}
where $\mathcal{Q}_\phi$ represents the nuclear charge related to boson scattering:
\begin{align}
    \mathcal{Q}_\phi = \sum_{N,q} g_\phi^{qS}\frac{m_N}{m_q} f_{T,q}^{(N)} \rightarrow g_\phi(14N+15.1Z)\, ,
\end{align}
for universal coupling $g_{Z'}$ and given proton and neutron numbers $Z$ and $N$, respectively.
The corresponding modified E$\nu$eS cross sections reads
\begin{align}\label{eq:cross_section_light_scalar_nu_e}
    \left(\frac{d\sigma}{dT_{e}}\right)_{{\rm E\nu eS} + \phi} = 
    \left(\frac{d\sigma}{dT}\right)_{{\rm E\nu eS}} + \frac{ (g_{\phi}^{\nu \mathrm{S}} g^{e \mathrm{S}}_{\phi})^{2}\,  m_{e}^{2} T_{e} }{4\pi E_{\nu}^{2} (2 m_{e} T_{e}  + m_{\phi}^{2})^{2}}\, ,
\end{align}
where we have already assumed universal coupling $g^{\nu \mathrm{S}}_{\phi}=g^{e \mathrm{S}}_{\phi}=g_{\phi}$.
Thus, for both interactions the scalar contribution can only enhance the expected number of signal events.


\subsection{The Weinberg angle at low energy}

One of the profound predictions of the SM is the connection of electromagnetic and weak gauge coupling or the electroweak gauge boson masses via the Weinberg angle 
\begin{align}
    \sin^{2}\theta_{W} = e^{2}/g^{2} = 1- m^{2}_{W}/m^{2}_{Z}\, .
\end{align} 
Since this relation is also respected by renormalization group (RG) running it tests the SM at quantum level.
Conversely, it can also be used to probe BSM physics as it is subject to quantum correction induced by new degrees of freedom.
Thus, any deviation from its SM expectation (at low energy) may point to the existence of new contributing (light) particles.
The most precise measurement of it was obtained by the LEP experiment ~\cite{ALEPH:2005ab} at the Z-boson resonance.
With the NuTeV measurement deviating from the Standard Model prediction by more than $3\sigma$~\cite{PhysRevLett.88.091802}, additional measurements with neutrinos to confirm a possible deviation from the prediction can prove very valuable. 
CE$\nu$NS allows to determine the Weinberg angle at even lower momentum, thus testing the running of the renormalization group to low scales. 
In particular, deviations seen in CE$\nu$NS could hint to the presence of new particles in kinetic regions far from the $Z$ pole. 
The first measurement of CE$\nu$NS through the \textit{KKL CAEN} dataset now allows its determination for the first time as done in the following. 


\section{Results on beyond the Standard Model physics from CONUS and CONUS+ data}
\label{sec:results}

In this section, the results for the experimental datasets collected at the experimental sites in Brokdorf and Leibstadt are presented, respectively.  
The phenomenological models have been selected to cover a wide range of BSM extensions and are discussed in terms of the corresponding datasets analyzed. 
The applied analysis framework relies on previous CE$\nu$NS investigations of the \textsc{Conus} collaboration and consequently incorporates all systematics and uncertainties of the experimental setup.


\subsection{Improved limits on the neutrino magnetic moment and millicharge} 

The dataset \textit{KBR Lynx}, obtained during the measurement campaign performed at KBR with the \textsc{Lynx} DAQ system, is used to improve the constraints on the neutrino electromagnetic properties. 
The data of Run-1 and Run-2 are combined with the data of Run-4 and Run-5. 
For the latter, a broader energy region of $2-20\,\mathrm{keV}$ (Mid \& High, cf.\ tables \ref{tab:energy_regions} and \ref{tab:exposures_cenns}) is used, where the range of $8-12\,\mathrm{keV}$ is excluded due to the cosmogenic activation lines of germanium.

The investigation on electromagnetic properties of the neutrino is performed with the analysis framework described in section~\ref{analysis_framework}. 
In figure \ref{fig:nmm_example} in the appendix, a single detector analysis for the neutrino magnetic moment analysis is shown to confirm the stability of the fit. Satisfying agreement with the data is demonstrated and no significant hints for additional events due to a non-zero NMM are found. Therefore, upper limits at 90\% C.L.\ are derived with a one-sided profile likelihood ratio test. 
As in \cite{CONUS:2022qbb}, the obtained value is converted into a NMC limit via eq.~\eqref{eq:nmc_limit_from_nmm}. 
The obtained limits determined from the dataset \textit{KBR Lynx} are:
\begin{align}
    \mu_\nu &< 5.18\cdot 10^{-11}\mu_\mathrm{B}\,,  &
    |q_\nu| &< 1.76\cdot  10^{-12} e_0\, .
\end{align}
These results mark a distinct improvement to the previous CONUS results of $\mu_\nu < 7.5\cdot 10^{-11}\mu_\mathrm{B}$ and $|q_\nu| < 3.3\cdot  10^{-12} e_0$, getting closer to the GEMMA results \cite{Beda:2013mta} of $\mu_\nu < 2.9\cdot 10^{-11}\mu_\mathrm{B}$ and $|q_\nu| < 1.5\cdot  10^{-12} e_0$.
A listing of results obtained from the combination of different datasets is given in appendix~\ref{app:nmm_nmc_limits}.


\begin{figure}[t]
    \centering
    \includegraphics[width=\linewidth]{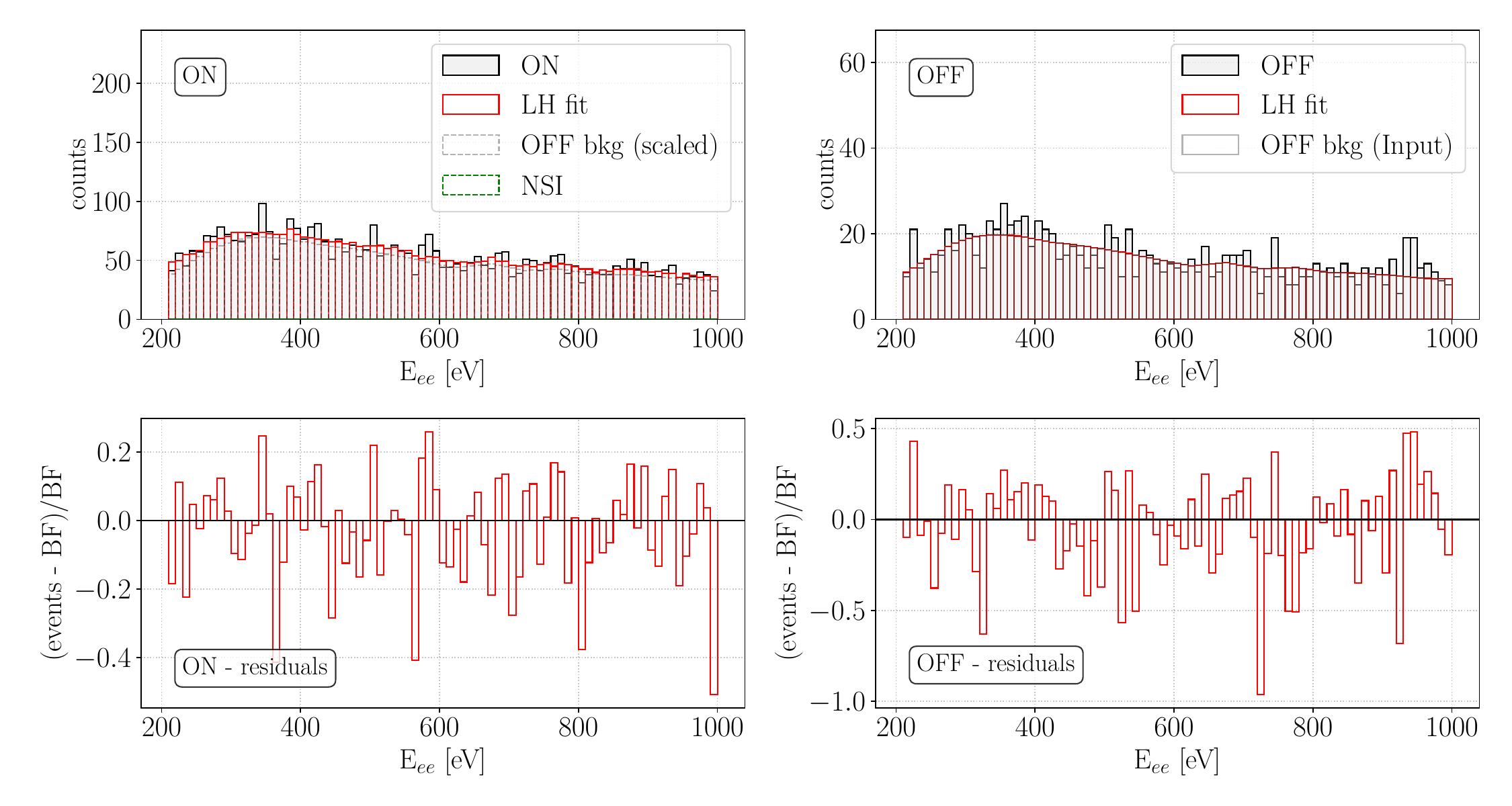}
    \caption{Exemplary fit for a single detector (C1) in the vector NSI analysis of the dataset \textit{KBR CAEN}. The best fit curve (red) and measured data (gray) are superimposed for both the ON (left) and OFF (right) data as well as the NSI signal contribution to the best fit (green) for the ON dataset. Corresponding residuals are shown in the bottom row.}
    \label{fig:nsi_example5}
\end{figure}

\begin{figure}[tb]
    \centering
    \includegraphics[width=0.6\linewidth]{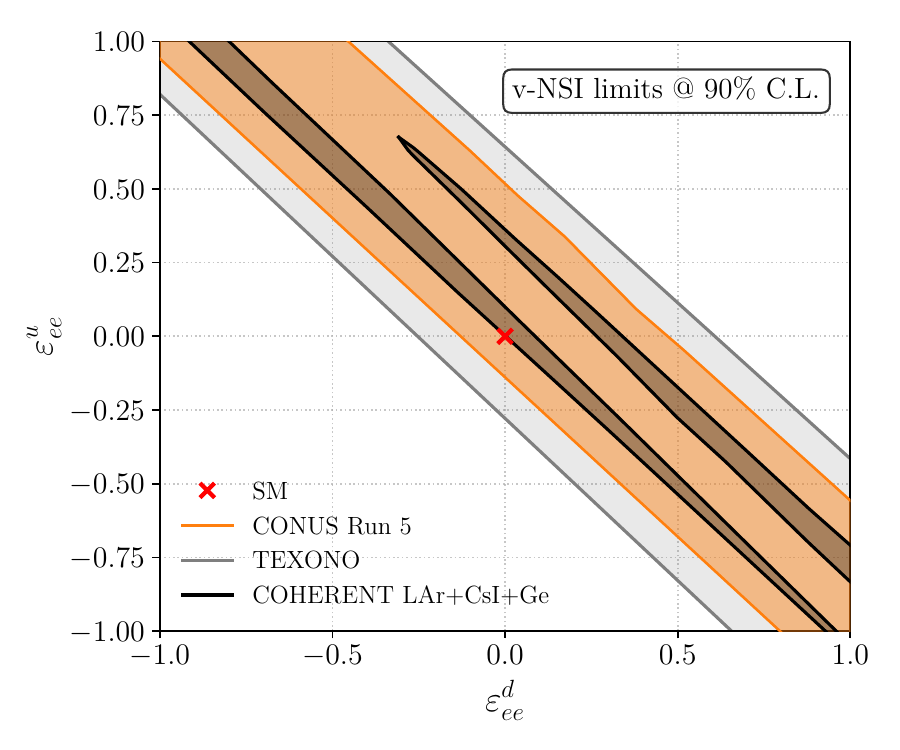}
    \caption{Limits on coupling of non-standard neutrino interaction of vector-type (at 90\% C.L.) from the dataset \textit{KBR CAEN} compared to existing limits determined via CE$\nu$NS. 
    The standard model (null coupling) is indicated with a red cross and the limits from the present analysis and further experiments (\textsc{Coherent}~\cite{Liao:2024qoe} and \textsc{Texono}~\cite{AtzoriCorona:2025ygn}) are shown in orange and gray, respectively.}
    \label{fig:nsi_limits5}
\end{figure}

\begin{figure}
    \centering
    \includegraphics[width=0.8\linewidth]{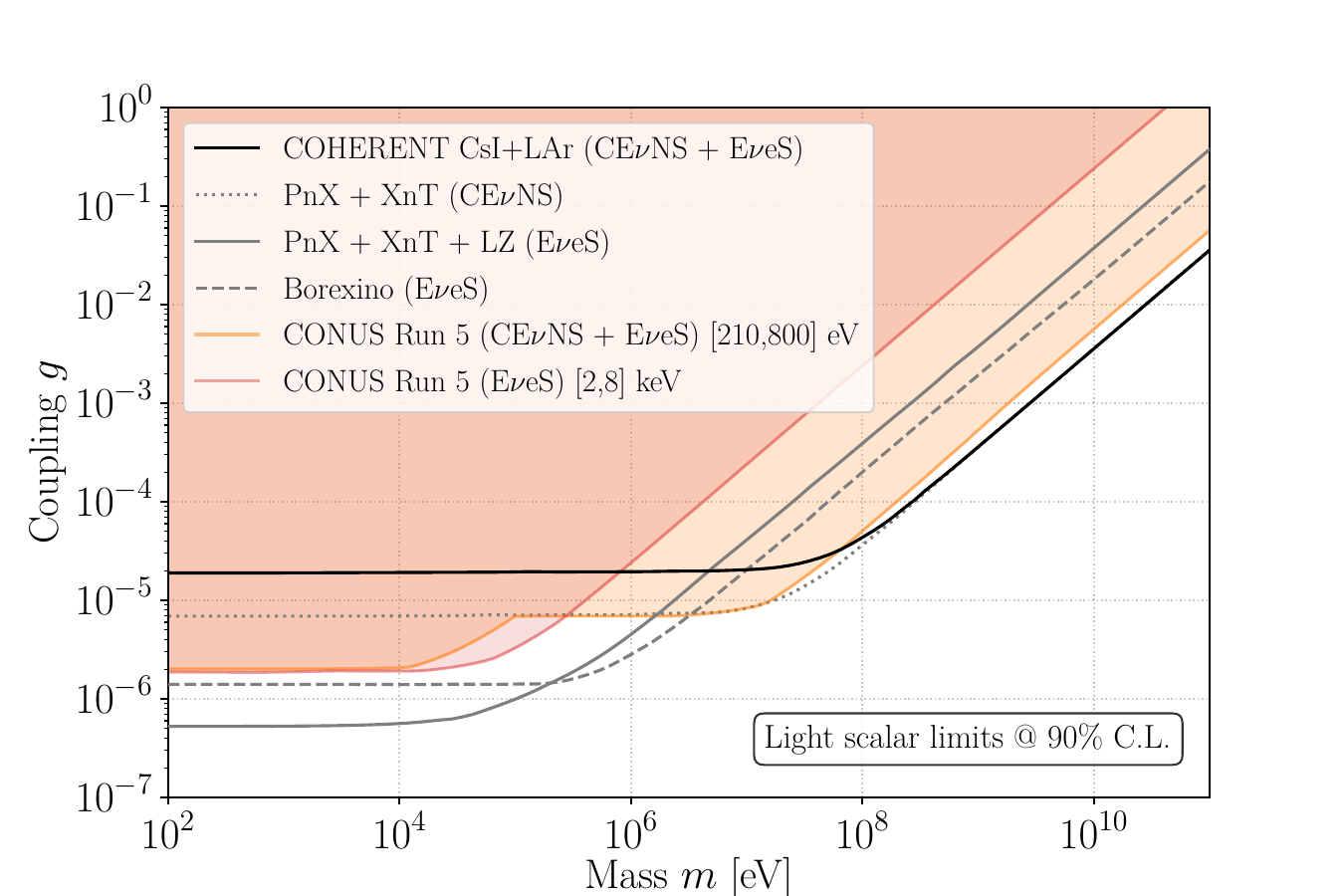}
    \caption{Limits on a light scalar mediator with universal coupling. Constraints on the parameter space of coupling and mediator mass are deduced at 90\% C.L.\ for the dataset \textit{KBR CAEN}. In orange, the analysis of the low energy ROI for both CE$\nu$NS and E$\nu$eS interaction channels and in red, the analysis of the mid energy ROI for E$\nu$eS are shown. Other competitive experiments in the CE$\nu$NS channel are shown in black (\textsc{Coherent} \cite{DeRomeri:2022twg}) and dotted gray (PnX + XnT \cite{DeRomeri:2024dbv}), in the E$\nu$eS channel in gray (PnX + XnT + LZ \cite{DeRomeri:2024dbv}) and dashed gray (Borexino \cite{DeRomeri:2024dbv}).}
    \label{fig:LS_limits5}
\end{figure}


\begin{figure}[t]
    \centering
    \includegraphics[width=0.8\linewidth]{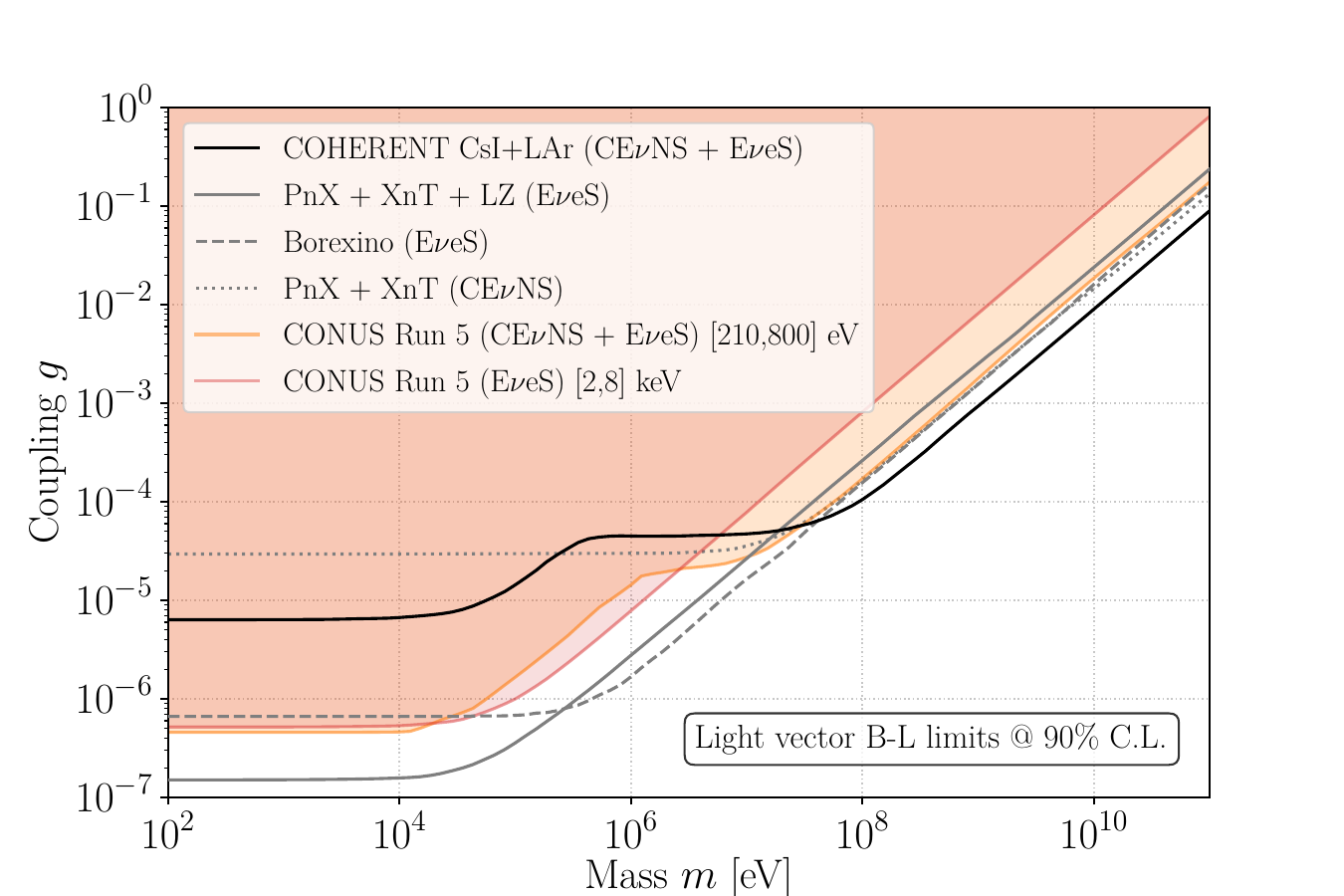}\\
    \includegraphics[trim=0 0 0 30,clip,width=0.8\linewidth]{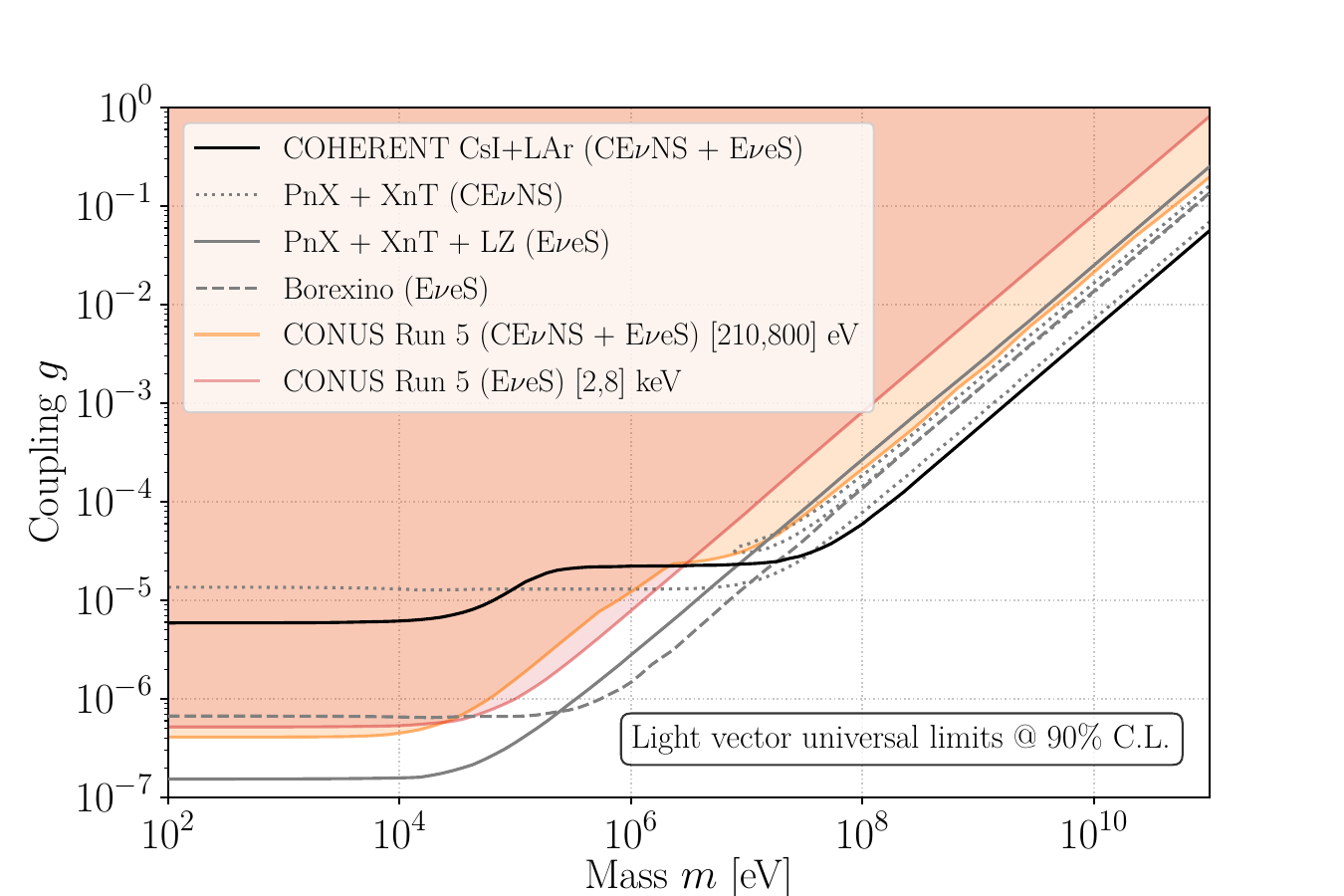}
    \caption{Top: Limits on a light new vector mediator (at 90\% C.L.) charged under $U(1)_{\mathrm{B-L}}$ from the dataset \textit{KBR CAEN}. Bottom: Limits on a light new vector mediator (at 90\% C.L.) with universal coupling from the dataset \textit{KBR CAEN}. In orange, the analysis of the low energy ROI for both CE$\nu$NS and E$\nu$eS interaction channels and in red, the analysis of the mid energy ROI for E$\nu$eS are shown. Other competitive experiments in the CE$\nu$NS channel are shown in black (\textsc{Coherent} \cite{DeRomeri:2022twg}) and dotted gray (PnX + XnT \cite{DeRomeri:2024dbv}), in the E$\nu$eS channel in gray (PnX + XnT + LZ \cite{DeRomeri:2024dbv}) and dashed gray (Borexino \cite{DeRomeri:2024dbv}).}
    \label{fig:LVB_limits5}
    \label{fig:LVu_limits5}
\end{figure}


\subsection{Updated constraints from the last dataset at the Brokdorf site}

Now, the last dataset collected at the KBR site (taken with the \textsc{CAEN} DAQ, see section \ref{sec:datasets}) is used to constrain vector NSI as well as light mediating particles. 
For the former, only the data in the energy region \textit{Low} are investigated with the CE$\nu$NS channel, while we use CE$\nu$NS and E$\nu$eS to probe the effect of new mediators in investigations of the energy regions \textit{Low} and \textit{Mid}, cf.\ table~\ref{tab:energy_regions}.

\paragraph{Non-standard interactions of vector-type} 
With the analysis framework previously described, limits at 90\% C.L.\ are obtained in the $\{\varepsilon^{uV}_{ee}, \varepsilon^{dV}_{ee}\}$ parameter plane. 
In figure \ref{fig:nsi_example5} we show a single detector analysis to confirm the stability of the fit and satisfying agreement with the data. 
As indicated, no significant BSM events are found.  
The results of the combined analysis of dataset \textit{KBR CAEN} are illustrated in figure~\ref{fig:nsi_limits5} together with other limits determined in the CE$\nu$NS context~\cite{Liao:2024qoe,AtzoriCorona:2025ygn}.%
A typical band-like structure emerges.
Indeed, from equation~\eqref{eq:weak_nsi}, we are only sensitive to the linear combination of the two parameters. 
There are two peculiar features to be noted about the \textsc{Coherent} limits.
Firstly, two different bands are defined. 
Once the CE$\nu$NS signal comes into reach, a new degeneracy of the NSI gains relevance, i.e.\ the presence of the squared of the weak charge in the cross section, leading to two different $\varepsilon$ configurations being capable of producing the same number of events and thus creating a double-band structure.
Secondly, the lines defining these two bands are slanted, thus cutting more of the parameter space. 
This originates in the combination of the three different target materials. 
Indeed, looking at the relation between the weak nuclear charge and the NSI parameters in equation \eqref{eq:weak_nsi}, it can be seen how the slope of this linear relation changes depending on the number of protons and neutrons in the nucleus. 
Thus, combining different target materials allows to constrain the parameters further.

With the dataset \textit{KBR CAEN} \textsc{Conus} is about to resolve the two-band structure as can be seen for the evolution of confidence levels for a fixed $\varepsilon$ given in appendix~\ref{fig:app_run5_figurelimits_vector_nsi_section}. 
As reported in section \ref{sec:bsm_topics}, setting a limit on the new couplings allows to constrain the scale of new physics.
With the limits of figure \ref{fig:nsi_limits5} this scale can be constrained to be $\Lambda_{\rm NSI}\geq 135 \ \rm{GeV}$, which improves previous \textsc{Conus} constraints~\cite{CONUS:2021dwh}.

\paragraph{Light scalar mediator}
Next, the parameter space of a universally coupled light scalar mediator is investigated by looking for effects on both CE$\nu$NS (energy regions \textit{Low} and \textit{Mid}) and E$\nu$eS (\textit{Mid}), respectively.
The limits at 90\% C.L. in the $\{m_\phi, g_\phi\}$ parameter space are shown in figure~\ref{fig:LS_limits5}. 
The plot shows a composite shape with kinks and plateaus. 
The compositeness comes from considering both the SM channels at the same time, whereas the kink structure comes from the shape of the cross section of eq.~\eqref{eq:cross_section_light_scalar_cenns}. 
Indeed, if $2 m_x T_x >> m_\phi^2$ we will only be sensitive to the coupling, giving an increase in sensitivity in the x axis.
In the other case, $2 m_x T_x << m_\phi^2$, we will be sensitive to $g_\phi/m_\phi$, thus giving a linear increase in sensitivity in the parameter space. 
Constraints from other CE$\nu$NS experiments~\cite{DeRomeri:2022twg,DeRomeri:2024dbv} (\textsc{Coherent} and PnX + XnT) and large scale dark matter direct detection experiments (Borexino and PnX + XnT + LZ) are given as well.
With the CE$\nu$NS channel, a sensitivity to couplings down to $7\cdot 10^{-6}$ was reached, placing the experiment on par with PnX + XnT at low mediator masses, but not quite reaching the sensitivity of \textsc{Coherent} for higher masses (since a higher neutrino energy shifts the location of the kink to the right). Further, this marks a 30\% increase in sensitivity compared to the analysis for the previous runs \cite{CONUS:2021dwh}. The E$\nu$eS channel allowed to probe couplings down to $2\cdot 10^{-6}$, where DMDD experiments still lead the global sensitivity.

\paragraph{Light vector mediator}

Corresponding limits for the two light vector models under study are given in figure~\ref{fig:LVB_limits5}.
Analogous structures of plateaus and linear rises to the light scalar case can be seen since the kinematics of the interaction are the same.

Especially for the $B-L$ coupling, \textsc{Conus} sets competitive limits for higher mediator masses where the CE$\nu$NS interaction dominates the cross section.

In the CE$\nu$NS interaction channel couplings down to $2 \cdot 10^{-4}$ for both universal and B-L couplings were probed, setting best limits for the channel at lower mediator masses for the B-L case while not being able to reach PnX + XnT's sensitivity for the universal one. At higher mediator masses, \textsc{Coherent} dominates the sensitivity here. The sensitivity for this channel saw an increase of 33\% compared to the previous analysis \cite{CONUS:2021dwh}.
In the E$\nu$eS interaction channel, couplings down to $4\cdot 10^{-7}$ were probed, while large DMDD experiments are still leading.



\begin{figure}[htb]
    \centering
    \includegraphics[width=\linewidth]{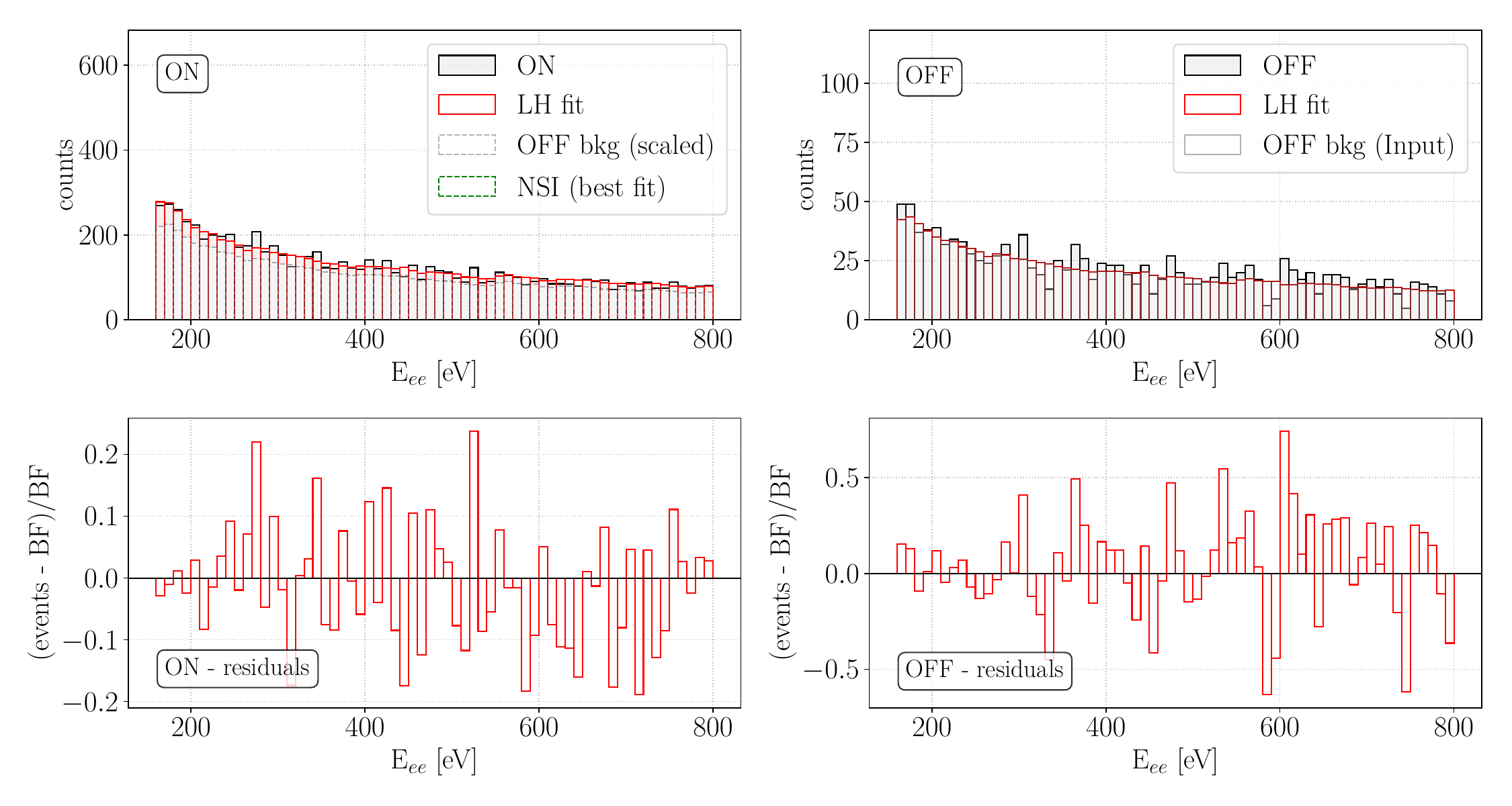}
    \caption{Exemplary fit for a single detector (C3) in the vector NSI analysis of the dataset \textit{KKL CAEN}. The best fit curve (red) and measured data (gray) are superimposed for both the ON (left) and OFF (right) data as well as the NSI signal contribution to the best fit (green) for the ON dataset. Corresponding residuals are shown in the bottom row.}
    \label{fig:nsi_example1}
\end{figure}

\begin{figure}[!htb]
    \centering
    \includegraphics[width=0.6\linewidth]{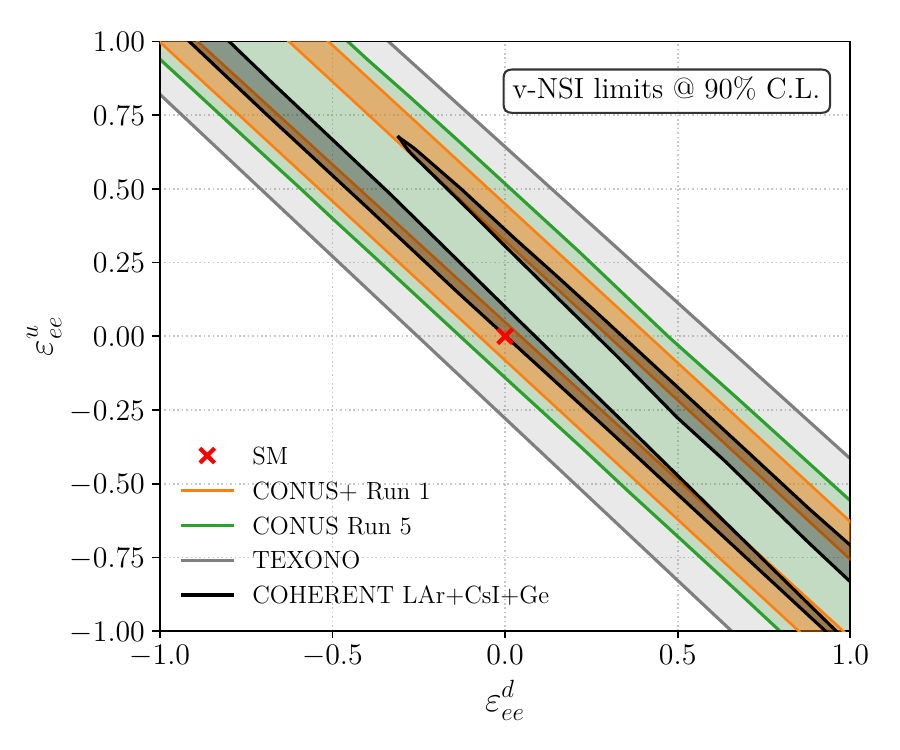}
    \caption{Limits on non-standard neutrino interaction of vector-type (at 90\% C.L.) from the dataset \textit{KKL CAEN} compared to existing limits determined via CE$\nu$NS. 
     At null coupling the standard model is shown as a red cross, the limits from the present analysis in orange and the ones from the previous analysis in green, further experiments in gray (\textsc{Coherent} \cite{Liao:2024qoe} and \textsc{Texono} \cite{AtzoriCorona:2025ygn}).}
    \label{fig:nsi_limitsplus}
\end{figure}

\begin{figure}[!htb]
    \centering
    \includegraphics[width=0.8\linewidth]{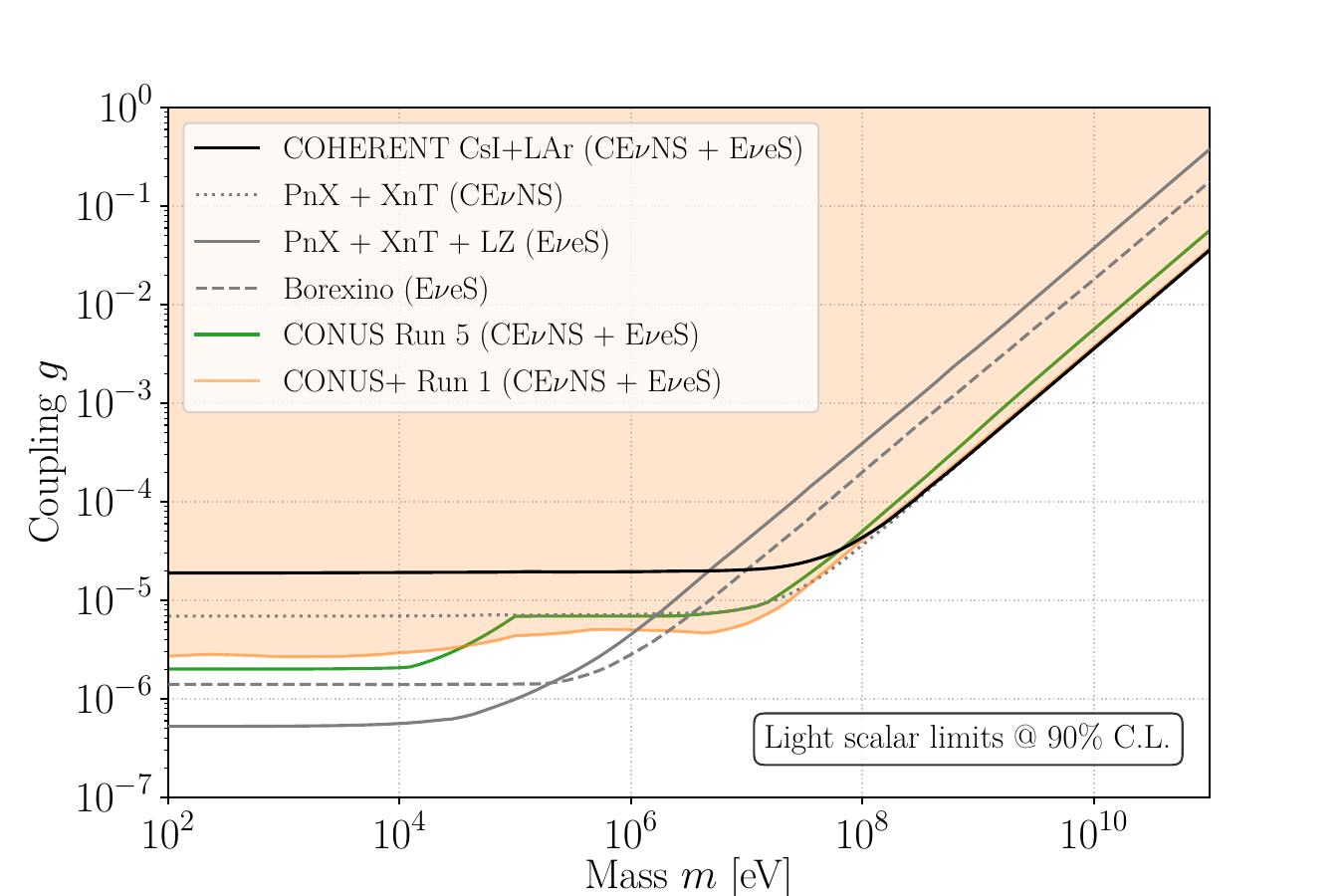}
    \caption{Limits on a light scalar mediator with universal coupling. Constraints on the parameter space of coupling and mediator mass are deduced at 90\% C.L.\ for the dataset \textit{KKL CAEN}. In orange, the analysis for both the low and mid energy ROIs for both CE$\nu$NS and E$\nu$eS interaction channels is shown. In green, the result from the previous analysis is displayed. Other competitive experiments in the CE$\nu$NS channel are shown in black (\textsc{Coherent} \cite{DeRomeri:2022twg}) and dotted gray (PnX + XnT \cite{DeRomeri:2024dbv}), in the E$\nu$eS channel in gray (PnX + XnT + LZ \cite{DeRomeri:2024dbv}) and dashed gray (Borexino \cite{DeRomeri:2024dbv}).}
    \label{fig:LS_limitsplus}
\end{figure}


\subsection{New physics limits after the first \texorpdfstring{CE$\nu$NS}{CEνNS} detection at the Leibstadt site}

Now the data from the KKL site, i.e.\ the dataset \textit{KKL CAEN}, are investigated for BSM physics.
While we again probe only the energy region \textit{Low} with CE$\nu$NS for any effects of vector NSI, we perform for the first time a investigations of light mediators in the combined energy region \textit{Low} and \textit{Mid} with both CE$\nu$NS and E$\nu$eS.

\paragraph{Non-standard interactions of vector-type} 
For the dataset \textit{KKL CAEN}, large improvements in terms of BSM parameters are expected as the signal is actually \textit{seen} in the data.
As for the previous analyses, for the vector NSI only the energy region \textit{Low} is considered.
An exemplary fit is reported in figure \ref{fig:nsi_example1}, showing again satisfying agreement with the data. 
In figure \ref{fig:nsi_limitsplus}, the 90\% C.L.\ limits on the vector NSI parameters deduced from the first KKL data (\textit{KKL CAEN}) are shown. 
The newly found CE$\nu$NS signal allowed to resolve the two band structure coming from the degeneracy with said signal.
To illustrate this improvement, we show the confidence levels in dependence of one of the NSI parameters with the other one fixed in appendix~\ref{fig:app_run1_limits_vector_nsi_section} for the datasets \textit{KBR CAEN} and \textit{KKL CAEN}, respectively.
These allow to visualize the emergence of the two bands and the improvements achieved in this data collection period. 
The overall sensitivity was also further improved compared to the final run at KBR, making the experiment competitive with the \textsc{Coherent} result. 
A further analysis combining the present result with \textsc{Coherent}'s could allow to cut a substantial part of the parameter space. 
From the newly constrained $\varepsilon$ parameters, the sensitivity to a new physics scale can be estimated, as described in section \ref{sec:bsm_topics}, and amounts to $145 \ \rm{GeV}$.


\begin{figure}[htb]
    \centering
    \includegraphics[width=0.8\linewidth]{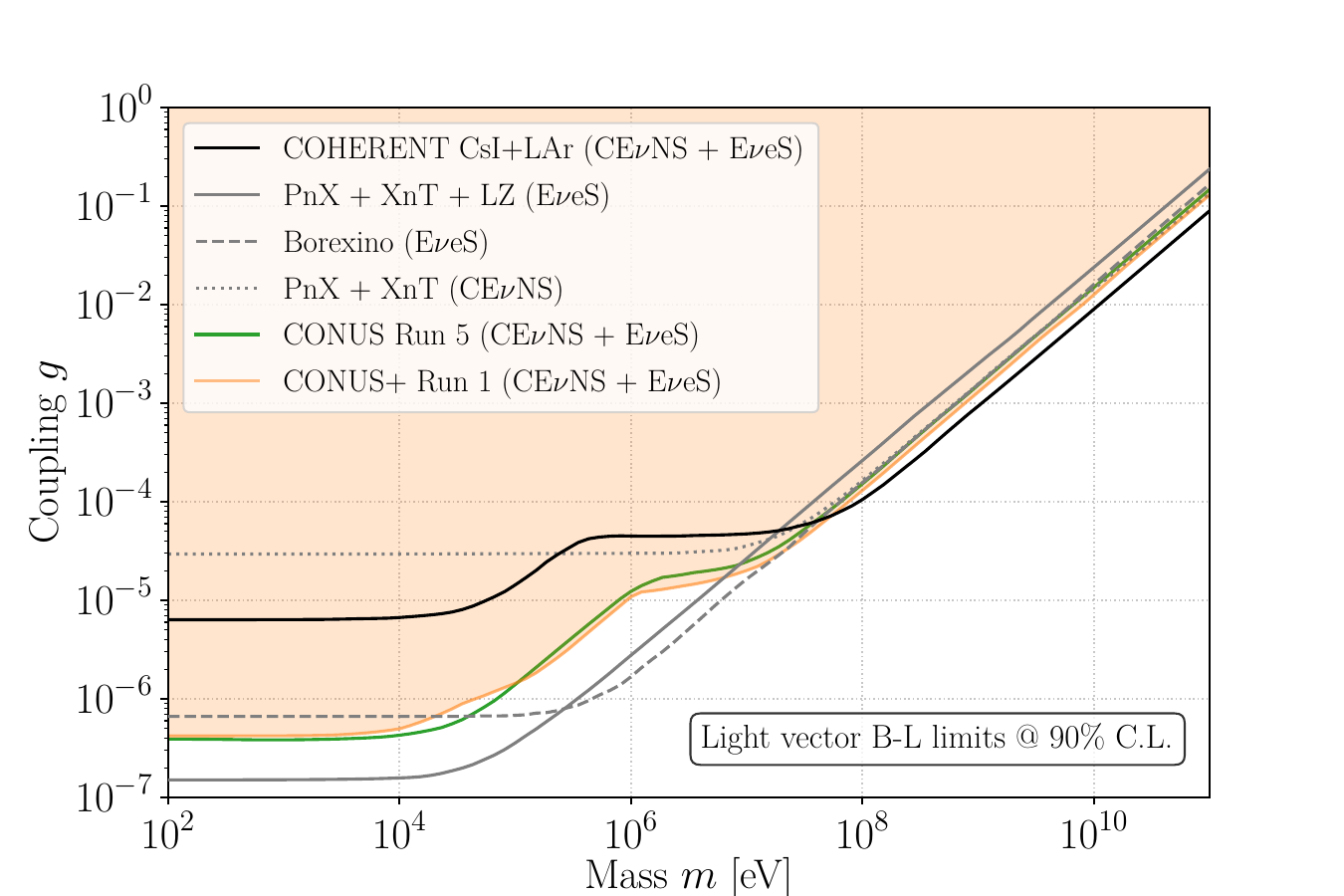}\\
    \includegraphics[trim=0 0 0 30,clip,width=0.8\linewidth]{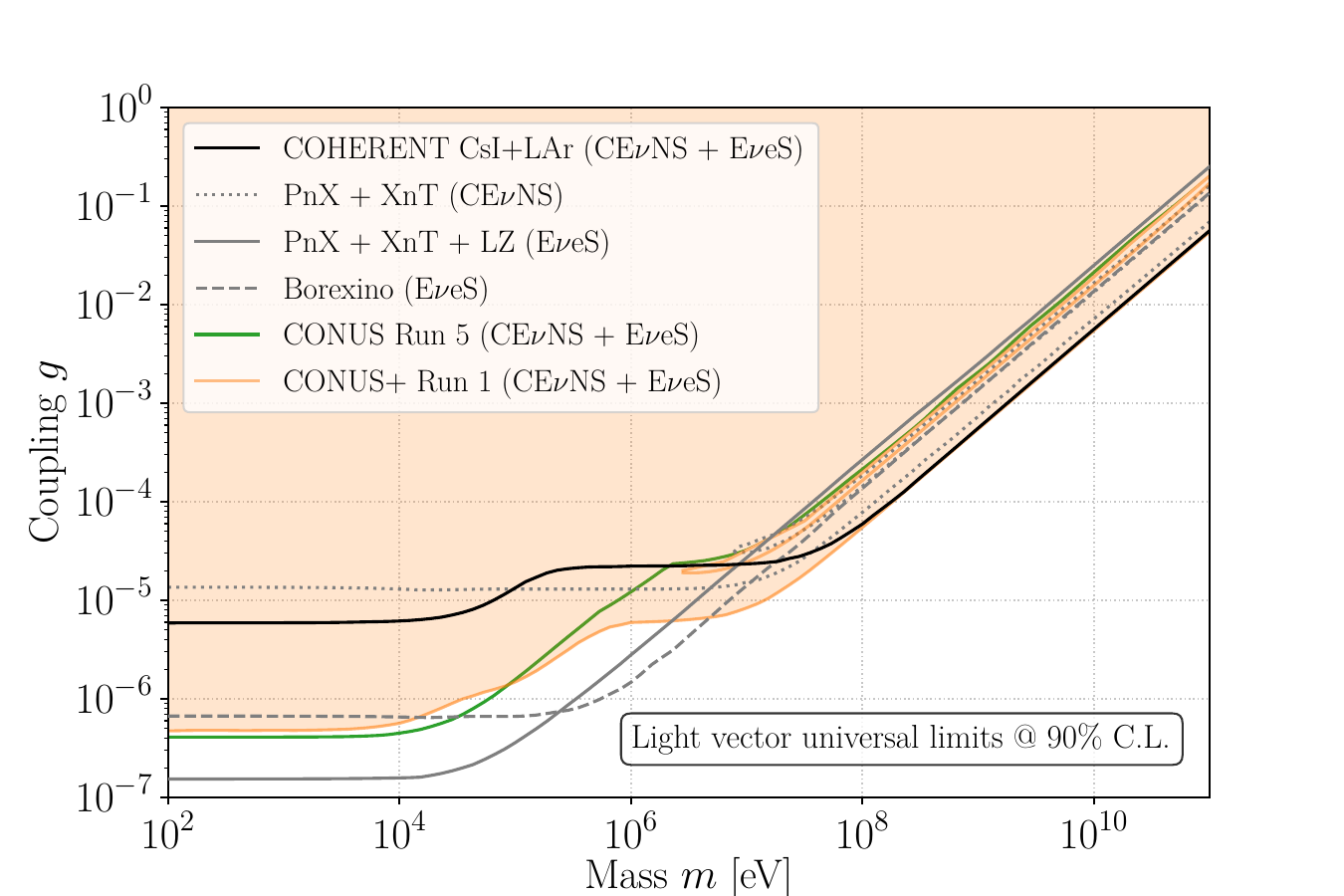}
    \caption{Top: Limits on a light new vector mediator (at 90\% C.L.) charged under $U(1)_{\mathrm{B-L}}$ from the dataset \textit{KKL CAEN}. Bottom: Limits on a light new vector mediator (at 90\% C.L.) with universal coupling from the dataset \textit{KKL CAEN}. In orange, the analysis for both the low and mid energy ROIs for both CE$\nu$NS and E$\nu$eS interaction channels is shown. In green, the results from the previous analysis are displayed. Other competitive experiments in the CE$\nu$NS channel are shown in black (\textsc{Coherent} \cite{DeRomeri:2022twg}) and dotted gray (PnX + XnT \cite{DeRomeri:2024dbv}), in the E$\nu$eS channel in gray (PnX + XnT + LZ \cite{DeRomeri:2024dbv}) and dashed gray (Borexino \cite{DeRomeri:2024dbv}).}
    \label{fig:LV_limitsplus}
\end{figure}


\paragraph{Light scalar mediator} 
Our results of the searches for a light scalar can be found in figure \ref{fig:LS_limitsplus}.
The 90\% C.L.\ limits exhibit the expected structure described for the previous run, but not in a combined way.
However, one can still infer the constant plateau regions for E$\nu$eS and CE$\nu$NS, while a linear region at high mediator masses is dominated by CE$\nu$NS.
Probing couplings down to $5\cdot 10^{-6}$ in this region, the experiment was able to set the best limits in the channel at lower mediator masses.
Moreover, in the region around the transition between the two regimes at $m\approx 10^7 \ \mathrm{eV}$ \textsc{Conus} provides better limits than other DMDD and coherent scattering experiments.
In the E$\nu$eS channel, couplings down to $3 \cdot 10^{-6}$ were probed while the DMDD experiments (Borexino and PnX + XnT + LZ) still provide the best constrains.
Comparing to the results of the previous analysis of the \textit{KBR CAEN} dataset, a marked improvement in the CE$\nu$NS dominated region can be seen, while in the E$\nu$eS dominated one no sensitivity increase is observed, which can be attributed to a higher overall background level (less overburden). 
\paragraph{Light vector mediator}
A similar behavior is observed in figure \ref{fig:LV_limitsplus} by the 90\% C.L. limits for the universally and $B-L$-coupled light vector mediators.
With couplings constrained down to $10^{-5}$ and $5 \cdot 10^{-6}$ in the CE$\nu$NS channel for the B-L and universal case, respectively, best limits are set improving by a margin on the ones already obtained for the previous run. 
For the universal light vector, the limits are on par with \textsc{Coherent} at high mediator masses, whereas they dominate at lower mediator masses. 
Probing down to coupling of $4 \cdot 10^{-7}$ with E$\nu$eS, the limits at low mediator masses are still around 60\% away from the leading constraint set by PnX + XnT + LZ and do not show any improvement from the previous run due to the higher background level. 
With the detection of the SM CE$\nu$NS signal, a new degeneracy in the universally coupled light vector rises from the potential destructive interference of the new signal. 
Indeed, at higher mediator masses, there is a thin elongated strip of non-exclusion that can be seen in figure \ref{fig:LV_limitsplus}. 
A better sensitivity with E$\nu$eS could help resolve this degeneracy if the non-exclusion strip were to be overtaken by the contour due to the electron scattering.

\begin{figure}[ht]
    \centering
    \includegraphics[width=.75\textwidth]{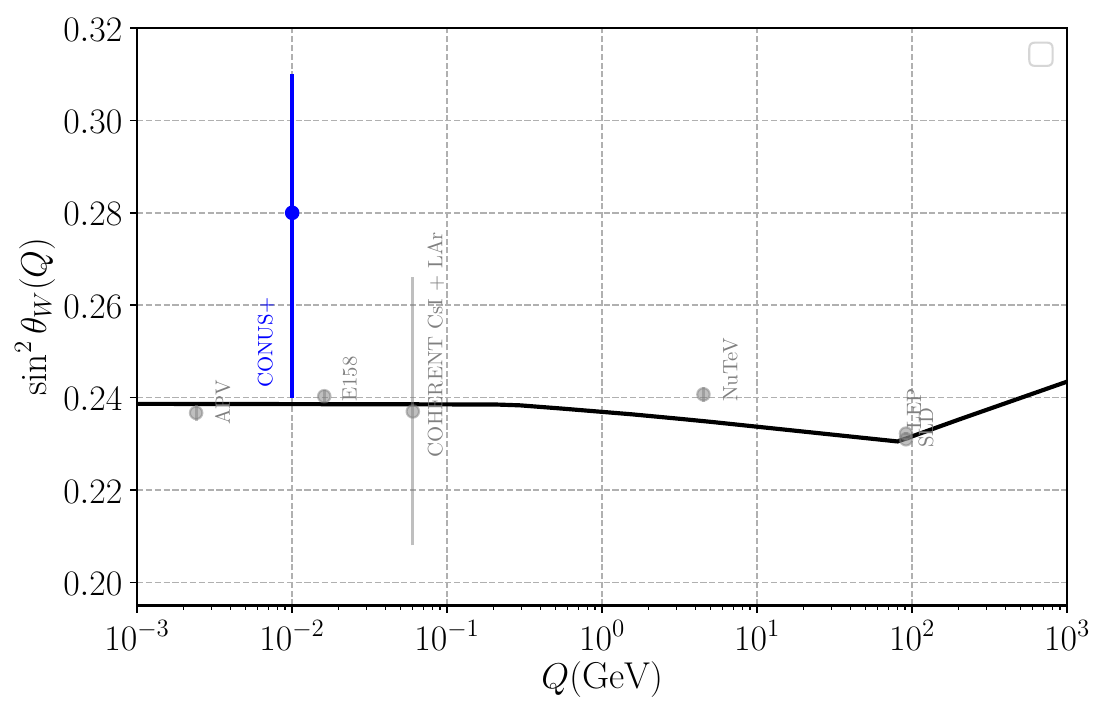}
    \label{fig:weinberg}
    \caption{The determined value of the Weinberg angle $\sin^{2}\theta_{W}$ is shown with $1\sigma$ uncertainty in blue. The gray data points indicate results from other experiments (\textsc{Apv} \cite{Wood:1997zq, Guena:2004sq}, E158 \cite{MOLLER:2014iki}, \textsc{Coherent} \cite{DeRomeri:2022twg}, NuTeV \cite{NuTeV:2001whx}, \textsc{Lep} \cite{ALEPH:2005ab} and \textsc{Sld} \cite{D0:2011baz}) at different momentum transfers. In black, the RG running within the SM expectation is indicated.}
\end{figure}

\paragraph{Weinberg angle at low energy}
A further way to test the SM is to constrain the Weinberg angle which appears in eq.~\eqref{eq:cross_section_sm_cenns}. 
Indeed, many BSM signatures can show up in the form of a modified Weinberg angle, making this quite interesting, and actually straightforward quantity to constrain when CE$\nu$NS is present in the data. 
The best fit with its $1\sigma$ error is shown in figure \ref{fig:weinberg} in comparison to other experiments constraining $\sin^{2}\theta_W$ at different momentum transfers. 
The resulting estimation from the present analysis is:
\begin{align}
    \sin^2{\theta_W} = 0.28^{+0.03}_{-0.04} \quad \text{(68 \% C.L.)} \quad ^{+0.06}_{-0.07} \quad \text{(90\% C.L.)}\, .
\end{align}
The result is about 1$\sigma$ away from the SM prediction. The result is achieved with similar precision to \textsc{Coherent} \cite{DeRomeri:2022twg} and is compatible with its result.


\begin{figure}[t]
    \centering
    \includegraphics[width=0.75\linewidth]{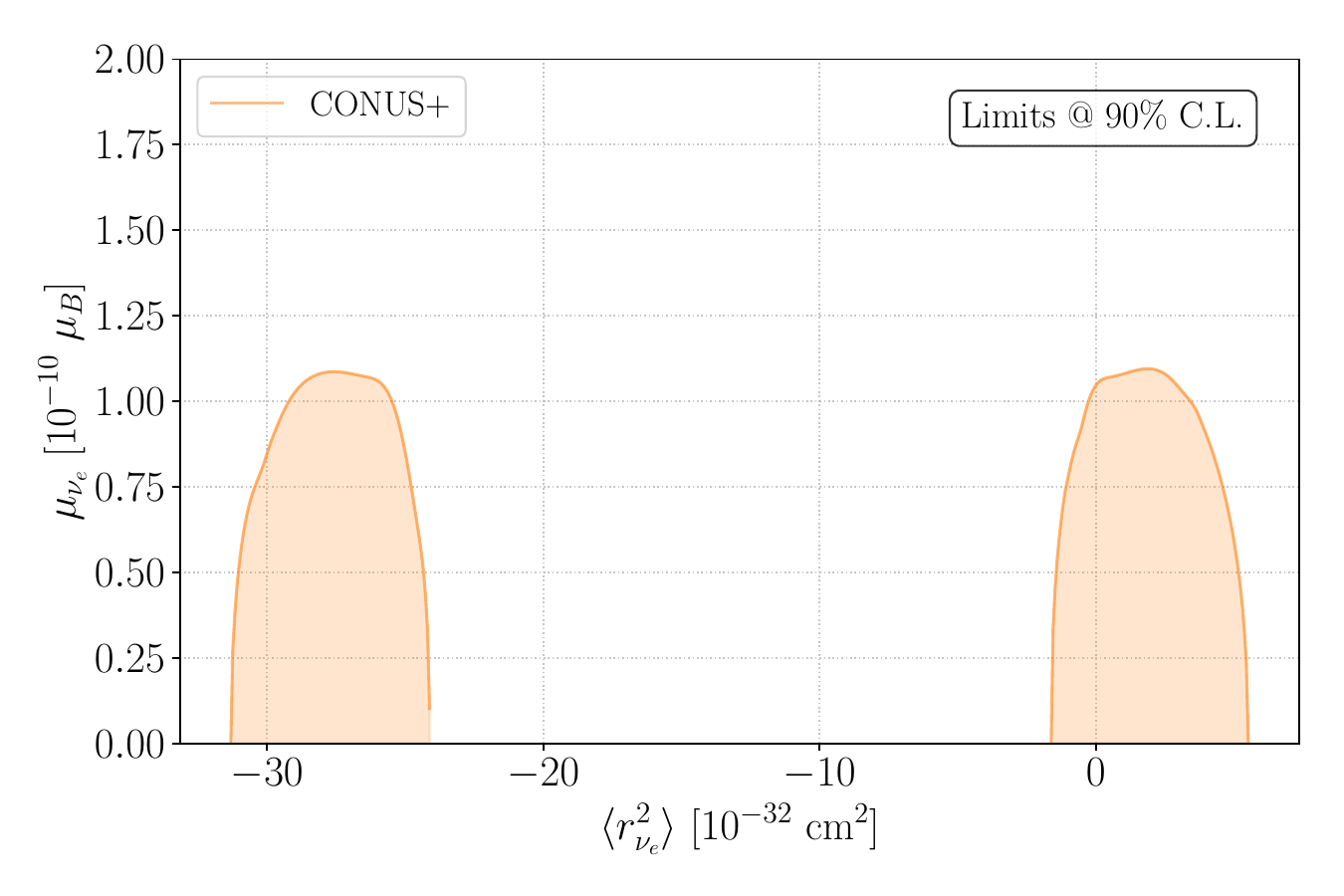}
    \caption{90\,\% C.L.\ contours obtained in the combined investigation of the NMM and the neutrino charge radius.}
    \label{fig:ncr}
\end{figure}


\paragraph{Neutrino electromagnetic properties}

Finally, we investigate the  combined energy region \textit{Low} and \textit{Mid} in terms of electromagnetic properties of the neutrino. 
Two independent analyses are carried out for the neutrino magnetic moment as well as the neutrino millicharge where both the E$\nu$eS and CE$\nu$NS interaction channels are considered simultaneously.
The resulting 90\,\% C.L.\ limits are:
\begin{align}
    & \mu_{\nu_e} < 1.09 \cdot 10^{-10} \mu_B\, , &
    |q_{\nu_e}| < 2.45 \cdot 10^{-12} e\, . 
\end{align}
The $q$ values and $(1-p)$ values showing the intersection with the 90\% threshold for both the magnetic moment and millicharge are displayed in figure \ref{fig:1_p_nmm} in the appendix.

%
As in the E$\nu$eS dominated light mediator cases, we do not expect to improve the sensitivity obtained in the previous run for the magnetic moment. 
Indeed, this analysis is dominated by the E$\nu$eS interaction. Further, the total exposure is substantially lower than the analysis of the \textit{KBR Lynx} dataset. 
Additionally, a narrower ROI is considered.

A different situation emerges for the NMC:
The obtained limit is closer to the one from the previous analysis.
Indeed, the $1/T^2$ dependence in eq.~\eqref{eq:nmc_cross_section} for the millicharge makes the cross section more sensitive to a lower energy threshold compared to the $1/T$ dependence in eq.~\eqref{eq:nmm_cross_section} for the neutrino magnetic moment.

Further, we allowed for a neutrino charge radius, cf.\ eq.~\eqref{eq:charge_radius}, and varying the NMM as well, a two-dimensional analysis in the $\{\langle r_{\nu_e}^2\rangle, \mu_{\nu_e}\}$ parameter space can be set up.
The corresponding limits (90\% C.L.) are reported in figure \ref{fig:ncr}. A degeneracy can be seen that originates from the modification of the Weinberg angle (see eq. \ref{eq:charge_radius}), which allows also for negative values of the charge radius.
The neutrino charge radius is restricted in the contour to being in $[-31.3, -24.1] \cup [-1.6, 5.5] \cdot 10^{-32} \ \rm{cm^2}$ and the NMM to the previously obtained limit. 
Compared to \textsc{Coherent}'s result \cite{DeRomeri:2022twg}, $\langle r_{\nu_e}^2\rangle \in [-61.2, -48.2] \cup [-4.7, 2.2] \cdot 10^{-32} \ \rm{cm^2}$, for the higher allowed region, in agreement with the trend seen for the Weinberg angle, the present result is higher, yet compatible.



\begin{table}[t]
    \centering
    \begin{tabular}{c c|c c}
    \hline
    \multicolumn{2}{c|}{results on}  & \textit{KBR Lynx/CAEN} & \textit{KKL CAEN} \\
    \hline
    \multicolumn{2}{c|}{$\mu_{\nu_e}$} & $<5.18 \cdot 10^{-11} \mu_B$ & $<1.09 \cdot 10^{-10} \mu_B$\\
    \multicolumn{2}{c|}{$q_{\nu_e}$} & $<1.76 \cdot 10^{-12} e$ & $<2.45 \cdot 10^{-12} e$\\
    \multicolumn{2}{c|}{$\sin^2{\theta_W}$} & $-$ & $0.28^{+0.03}_{-0.04}$\\
    \multicolumn{2}{c|}{NSI new physics scale (GeV)} & 135 & 145\\
    \hline
    \multirow{2}{*}{Light scalar} & $m_{\phi} \approx 1 \ \rm{keV}$ & $<2.0\cdot 10^{-6}$ & $<2.7\cdot 10^{-6}$\\
     & $m_{\phi} \approx 1 \ \rm{GeV}$ & $<5.3\cdot 10^{-4}$ & $<3.8\cdot 10^{-4}$\\
     \hline
    \multirow{2}{*}{Light vector universal} & $m_{Z'} \approx 1 \ \rm{keV}$ & $<4.1\cdot 10^{-7}$ & $<4.8\cdot 10^{-7}$\\
     & $m_{Z'} \approx 1 \ \rm{GeV}$ & $<2\cdot 10^{-3}$ & $<5\cdot 10^{-4}$\\
    \hline
    \multirow{2}{*}{Light vector B-L} & $m_{Z'} \approx 1 \ \rm{keV}$ & $<3.8\cdot 10^{-7}$ & $<4.2\cdot 10^{-7}$\\
     & $m_{Z'} \approx 1 \ \rm{GeV}$ & $<1.5\cdot 10^{-3}$ & $<1.3\cdot 10^{-3}$\\
    \end{tabular}
    \caption{Summary of main results obtained in the investigations of this work given for the individual datasets analyzed}
    \label{tab:results_conclusion}
\end{table}


\section{Conclusion}
\label{sec:conclusions}

Coherent elastic neutrino-nucleus scattering (CE$\nu$NS) has proven to be a useful tool for various investigations within and beyond the standard model.
Especially now, where we have successful detections from multiple neutrino sources, i.e.\ nuclear reactors, $\pi$DAR sources and the sun, phenomenological investigations may profit from the characteristics of the individual experiments and (anti-)neutrino emission.   
The \textsc{Conus} experiment aimed for the first CE$\nu$NS detection with high-purity Germanium detections in an advanced shield design in the vicinity of two nuclear reactors, the commercial power plants in Brokdorf (KBR), Germany from 2018-2022 and Leibstadt (KKL), Switzerland from 2023, cf.\ table~\ref{tab:site_properties} for site characteristics.
Experimental refinement before the transition to the latter, in particular the lowering of the detection threshold down to $\sim160$\,eV, allowed the first successful detection of CE$\nu$NS with reactor antineutrinos in 2025.
This article covers a set of yet unpublished investigations of data collected in the last experimental runs at KBR as well as BSM analyses of the first data gathered at KKL.
An overview of the analyzed data and their characteristics is given in tables \ref{tab:energy_regions} and \ref{tab:exposures_cenns}.
The performed studies use the analysis framework of previous SM and BSM CE$\nu$NS investigations and therefore incorporate the full experimental systematics that are summarized in table~\ref{tab:likelihood_uncertainty}.

With the first KKL dataset, the Weinberg angle $\sin^{2}\theta$ is constrained to $\sin^2{\theta_W} = 0.28^{+0.03}_{-0.04}$ in agreement with the SM prediction.
Further, previous \textsc{Conus} limits on the neutrino magnetic moment (NMM) and neutrino millicharge (NMC) are improved with an extended dataset collected at KBR to $\mu_{\nu_e} < 5.18 \cdot 10^{-11} \mu_B$ for the NMM and $q_{\nu_e} < 1.76 \cdot 10^{-12} e$ for the NMC.
In terms of vector-type non-standard interaction (NSI), the experiment resolves for the first time the characteristic double-band structure and constraints the corresponding new physics scale to be above $\Lambda_{\rm NSI}>145\,\rm{GeV}$.
Investigations of light mediator models improve existing \textsc{Conus} constraints and now probe couplings down to $2 \cdot 10^{-6}$ and $4 \cdot 10^{-7}$ for a universally coupled light scalar and a light (universally-coupled and $B-L$-charged) vector boson, respectively.
The main results of this work put the experiment in a competitive position within the field's effort to find hints of new physics and are reported in table~\ref{tab:results_conclusion} for both reactor sites. 
The collected data have proven valuable for testing various broad new physics classes, and may be used for further testing in the future. 
In the meantime, the \textsc{Conus+} experiment continues data taking at the KKL site, now with a further improved detector setup, an upgraded total active mass of $8.2\,\rm{kg}$ and pulse shape information of recorded signal events.
With next experimental milestone in reach, i.e.\ the next release of KKL data with even higher statistics, results of this work are expected to improve further.
As existing experiments transition to precision CE$\nu$NS measurements and many experiments especially at reactor sites are catching up, interesting new results for SM and BSM physics are expected in the near future, especially when information of all these experiments are combined.


\appendix
\section{Details of the statistical analyses}


\subsection{Exemplary fit for the neutrino magnetic moment investigation}

\begin{figure}[H]
    \centering
    \includegraphics[width=\linewidth]{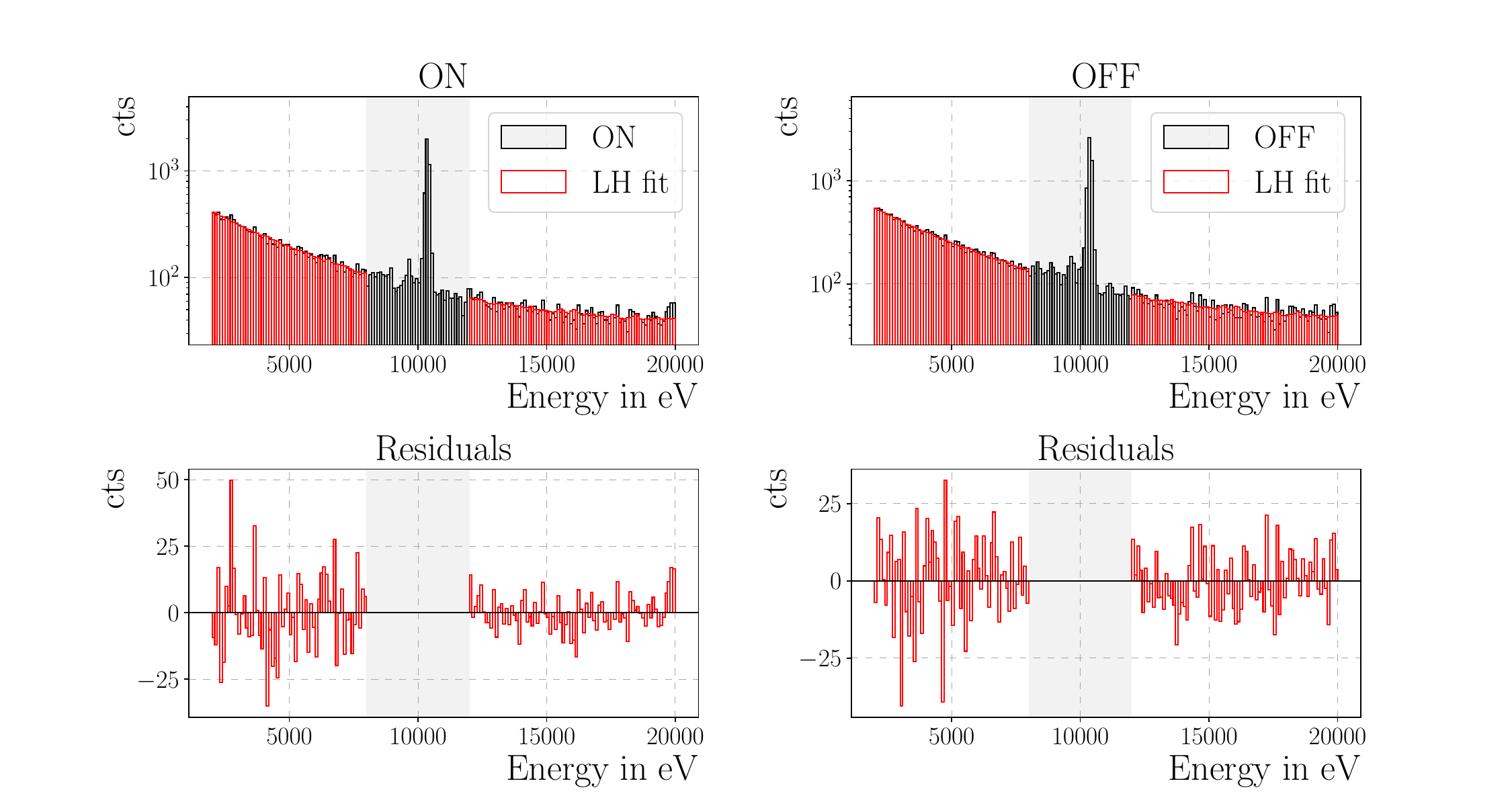}
    \caption{Exemplary fit for a single detector (C1) in the neutrino magnetic moment analysis of the dataset \textit{KBR Lynx}. The best fit curve (red) and measured data (black) are superimposed for both the ON (left) and OFF (right) data. The excluded range is shown in gray. Corresponding residuals are shown in the bottom row.}
    \label{fig:nmm_example}
\end{figure}

\subsection{Limits on neutrino electromagnetic properties for different datasets}\label{app:nmm_nmc_limits}


\begin{table}[H]
    \centering
    \begin{tabular}[h]{l|r|r}
    \hline
    Combined Runs & NMM (90\% C.L.) & NMC (90\% C.L.)\\
    \hline
    Run-1, Run-2 & $7.5 \cdot 10^{-11}\,\mathrm{\mu_B}$ & $3.3 \cdot 10^{-12}\,e_\mathrm{0}$\\
    Run-4, Run-5 & $5.57 \cdot 10^{-11}\,\mathrm{\mu_B}$ & $2.07 \cdot 10^{-12}\,e_\mathrm{0}$\\
    Run-1, Run-2, Run-4, Run-5 & $5.18 \cdot 10^{-11}\,\mathrm{\mu_B}$ & $1.76 \cdot 10^{-12}\,e_\mathrm{0}$\\
    \end{tabular}
    \caption[Upper limits on neutrino magnetic moment and millicharge for combined datasets with \textit{KBR Lynx} data]{Upper limits on neutrino magnetic moment and millicharge for Run-1 and Run-2 \cite{CONUS:2022qbb}, Run-4 and Run-5 and combined datasets with \textit{KBR Lynx} data. The upper limits are given at 90\% C.L.}
    \label{tab:nmm_kbr}
\end{table}


\subsection{Confidence levels for NSI parameters}


\begin{figure}[H]
\centering
\includegraphics[width=.85\textwidth]{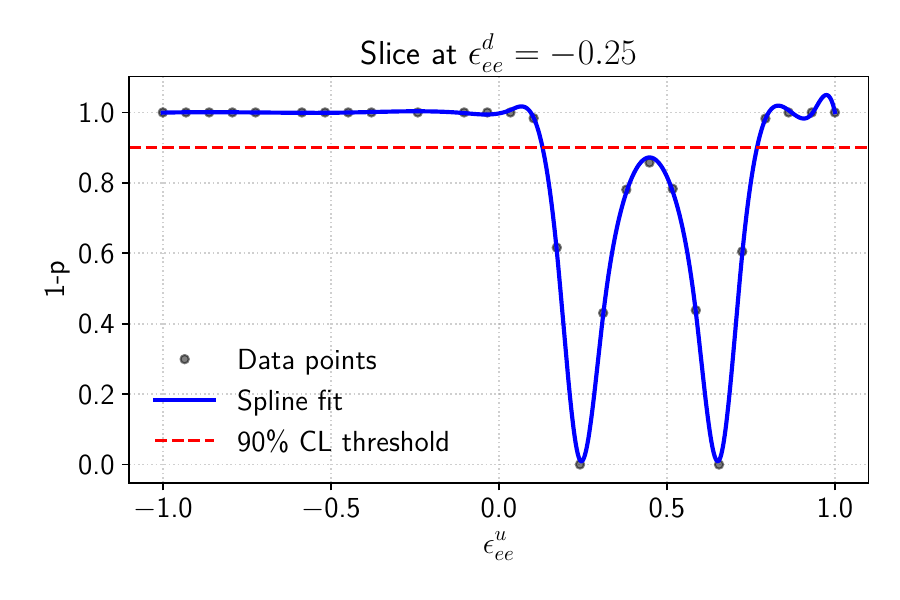}
\label{fig:app_run5_figurelimits_vector_nsi_section}
\caption{$(1-p)$ values for one parameter fixed ($\epsilon_{ee}^d=-0.25$) and intersection at 90\% C.L. for the KBR CAEN analysis.}
\end{figure}


\begin{figure}[H]
\centering
\includegraphics[width=.85\textwidth]{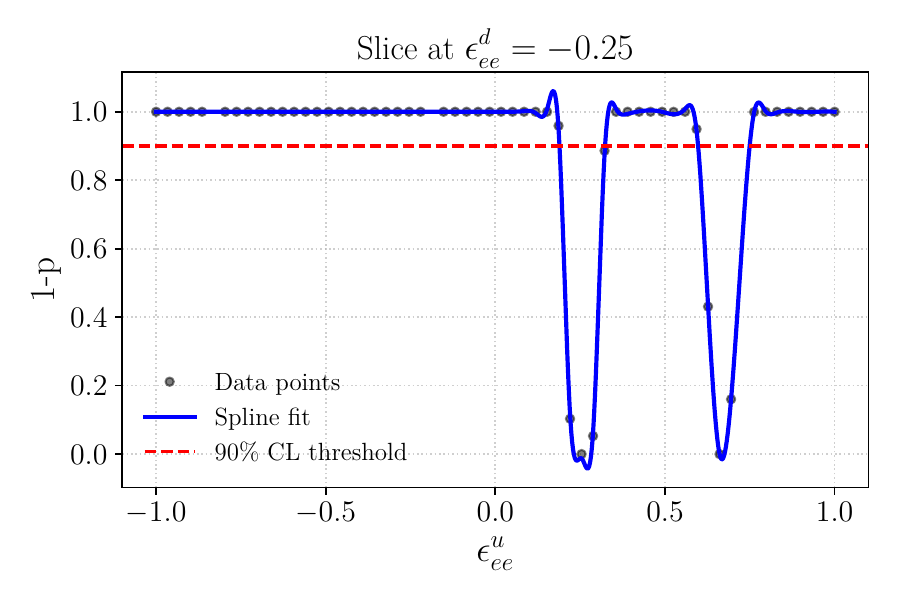}
\label{fig:app_run1_limits_vector_nsi_section}
\caption{$(1-p)$ values for one parameter fixed ($\epsilon_{ee}^d=-0.25$) and intersection at 90\% C.L. for the KKL CAEN analysis.}
\end{figure}


\subsection{Confidence levels for neutrino electromagnetic properties}

\begin{figure}[htb]
    \centering
    \includegraphics[width=0.85\linewidth]{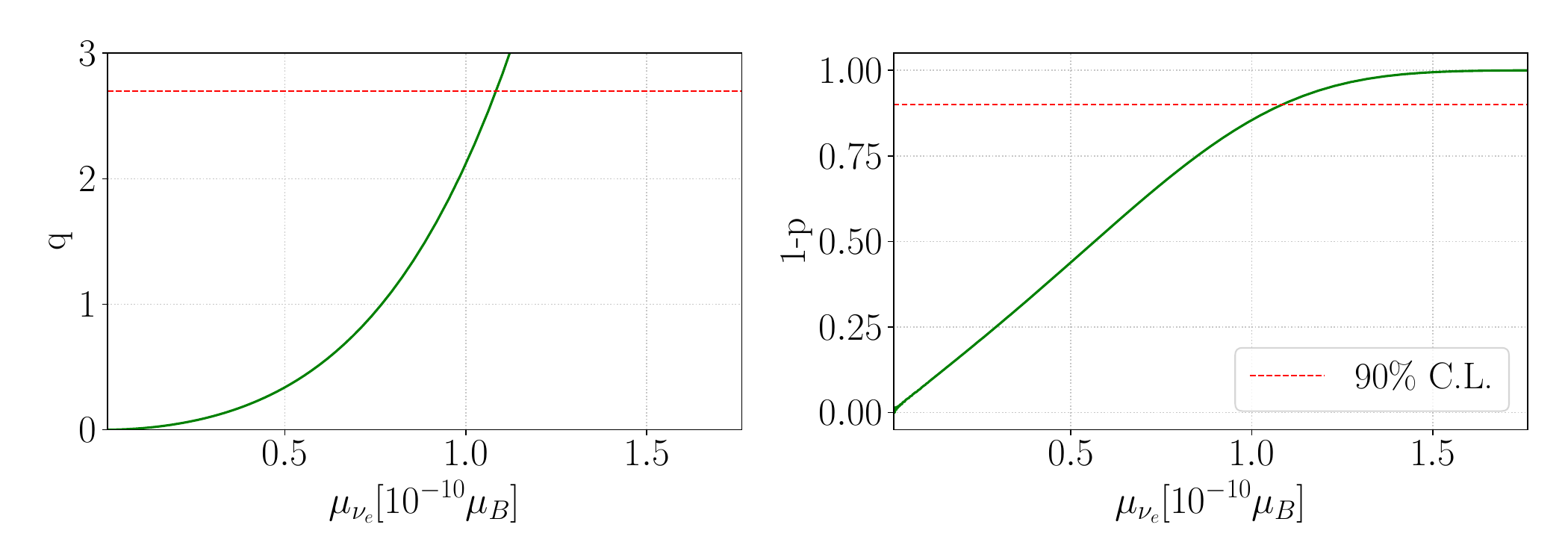}\\
    \includegraphics[width=0.85\linewidth]{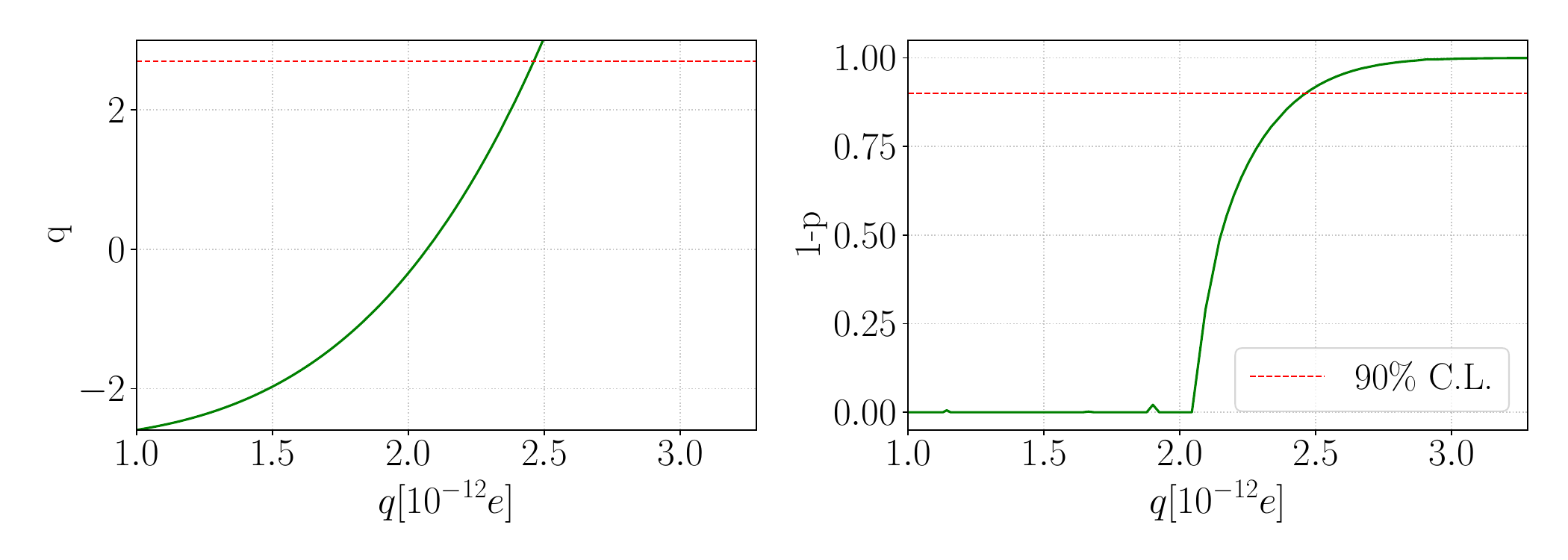}
    \caption{Top: $q$ values and $(1-p)$ values for the neutrino magnetic moment analysis, showing the intersection for the 90\% C.L. limit. Bottom: $q$ values and $(1-p)$ values for the neutrino millicharge analysis, showing the intersection for the 90\% C.L. limit.}
    \label{fig:1_p_nmm}
\end{figure}

\begin{figure}[H]
\centering
\includegraphics[width=.8\textwidth]{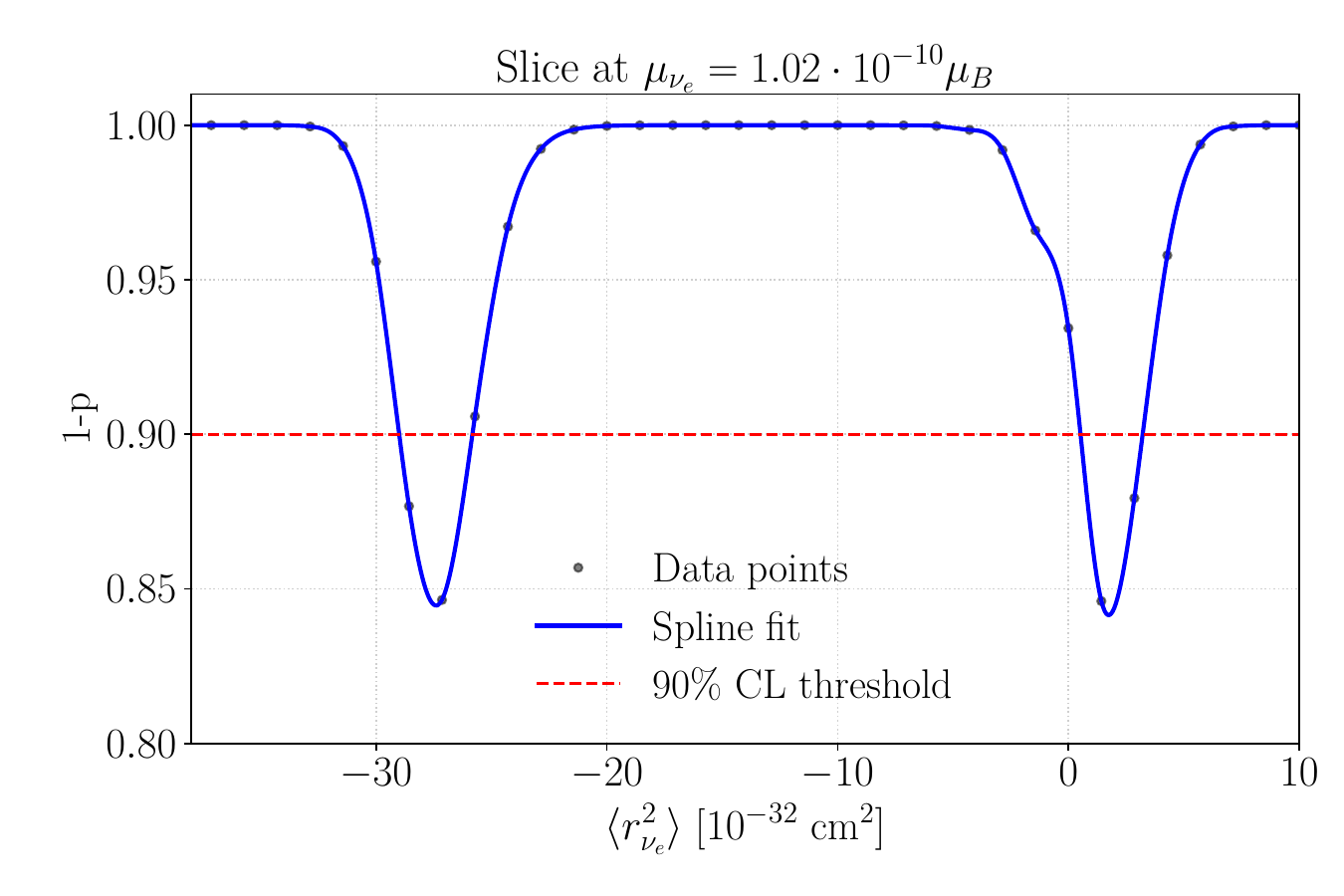}
\label{fig:app_run1_limits_ncr_section}
\caption{$(1-p)$ values for one parameter fixed ($\mu_{\nu_e}=1.02 \cdot 10^{-10} \mu_B$) and intersection at 90\% C.L.}
\end{figure}


\acknowledgments

We thank all divisions and workshops involved at the Max-Planck-Institut für Kernphysik in Heidelberg to set up the CONUS+ experiment, in particular T.\ Apfel, M.\ Reissfelder, T.\ Frydlewicz and J.\ Schreiner. 
The authors also thank Mirion Technologies (Canberra) in Lingolsheim for the detector upgrades and their highly professional support. 
We express our deepest gratitude to the PreussenElektra GmbH and Kernkraftwerk Leibstadt AG for great support and for hosting the \textsc{Conus} experiment, with special thanks to P.\ Graf, P.\ Kaiser, L.\ Baumann, R.\ Meili and A.\ Ritter. 



\bibliographystyle{JHEP}
\bibliography{literature}


\end{document}